\documentclass[runningheads]{cl2emult}

\usepackage{makeidx}   
\usepackage{graphicx}  

\begin{document}
\title*{Radiation Damping and Decoherence $\qquad \qquad$ in Quantum Electrodynamics}
\toctitle{Radiation Damping and Decoherence
 in Quantum Electrodynamics}
%
%
\titlerunning{Radiation Damping and Decoherence in Quantum Electrodynamics}
%

\author{Heinz--Peter Breuer\inst{1}
\and Francesco Petruccione\inst{1,2}}
\authorrunning{Heinz--Peter Breuer and Francesco Petruccione}
%
%
\institute{Fakult\"at f\"ur Physik, Albert-Ludwigs-Universit\"{a}t
Freiburg, Hermann--Herder--Str. 3, D--79104 Freiburg i. Br.,
Germany \and Istituto Italiano per gli Studi Filosofici, Palazzo
Serra di Cassano, Via Monte di Dio 14, I--80132 Napoli, Italy}

\maketitle              

\begin{abstract}
The processes of radiation damping and decoherence in Quantum
Electrodynamics are studied from an open system's point of view.
Employing functional techniques of field theory, the degrees of
freedom of the radiation field are eliminated to obtain the
influence phase functional which describes the reduced dynamics of
the matter variables. The general theory is applied to the
dynamics of a single electron in the radiation field. From a study
of the wave packet dynamics a quantitative measure for the degree
of decoherence, the decoherence function, is deduced. The latter
is shown to describe the emergence of decoherence through the
emission of bremsstrahlung caused by the relative motion of
interfering wave packets. It is argued that this mechanism is the
most fundamental process in Quantum Electrodynamics leading to the
destruction of coherence, since it dominates for short times and
because it is at work even in the electromagnetic field vacuum at
zero temperature. It turns out that decoherence trough
bremsstrahlung is very small for single electrons but extremely
large for superpositions of many-particle states.
\end{abstract}

\section{Introduction}
Decoherence may be defined as the (partial) destruction of quantum
coherence through the interaction of a quantum mechanical system
with its surroundings. In the theoretical analysis decoherence can
be studied with the help of simple microscopic models which
describe, for example, the interaction of a quantum mechanical
system with a collection of an infinite number of harmonic
oscillators, representing the environmental degrees of freedom
\cite{FEYNMAN-VERNON,LEGGETT}. In an open system's approach to
decoherence one derives dynamic equations for the reduced density
matrix \cite{GARDINER} which yields the state of the system of
interest as it is obtained from an average over the degrees of
freedom of the environment and the resulting loss of information
on the entangled state of the combined total system. The strong
suppression of coherence can then be explained 
by showing that the reduced density matrix equation leads to an 
extremely rapid transitions of a coherent superposition to an incoherent
statistical mixture \cite{ZUREK91,GIULINI}. For certain superpositions
the associated decoherence time scale is often found to be smaller than the
corresponding relaxation or damping time by many orders of
magnitude. This is a signature for the fundamental distinction
between the notions of decoherence and of dissipation. A series of
interesting experimental investigations of decoherence have been
performed as, for example, experiments on Schr\"odinger cat states
of a cavity field mode \cite{HAROCHE} and on single trapped ions
in a controllable environment \cite{WINELAND}.

If one considers the coherence of charged matter, it is the
electromagnetic field which plays the r\^{o}le of the environment.
It is the purpose of this paper to study the emergence of
decoherence processes in Quantum Electrodynamics (QED) from an
open system's point of view, that is by an elimination of the
degrees of freedom of the radiation field. An appropriate
technique to achieve this goal is the use of functional methods
from field theory. In section 2 we combine these methods with a
super-operator approach to derive an exact, relativistic
representation for the reduced density matrix of the matter
degrees of freedom. This representation involves an influence
phase functional that completely describes the influence of the
electromagnetic radiation field on the matter dynamics. The
influence phase functional may be viewed as a super-operator
representation of the Feynman-Vernon influence phase
\cite{FEYNMAN-VERNON} which is usually obtained with the help of
path integral techniques.

In section 3 we treat the problem of a single electron in the
radiation field within the non-relativistic approximation. Starting from
the influence phase functional, we formulate the reduced electron motion 
in terms of a path integral which involves an effective action functional. 
The corresponding classical equations of motion are demonstrated 
to yield the Abraham-Lorentz equation describing the radiation damping of 
the electron motion. In addition, the influence phase is shown to lead to a 
decoherence function which provides a measure for the degree of decoherence.

The general theory will be illustrated with the help of two
examples, namely a free electron (section 4) and an electron
moving in a harmonic potential (section 5). For both cases an
analytical expression for the decoherence function is found, which
describes how the radiation field affects the electron coherence.

We shall use the obtained expressions to investigate in detail the
time-evolution of Gaussian wave packets. We study the influence of
the radiation field on the interference pattern which results from
the collision of two moving wave packets of a coherent
superposition. It turns out that the basic mechanism leading to
the decoherence of matter waves is the emission of bremsstrahlung
through the moving wave packets. The resultant picture of
decoherence is shown to yield expressions for the decoherence time
and length scales which differ substantially from the conventional
estimates derived from the prominent Caldeira-Leggett master
equation. In particular, it will be shown that a superposition of
two wave packets with zero velocity does not decohere and, thus,
the usual picture of decoherence as a decay of the off-diagonal
peaks in the corresponding density matrix does not apply to
decoherence through bremsstrahlung.

We investigate in section 6 the possibility of the destruction of
coherence of the superposition of many-particle states. It will be
argued that, while the decoherence effect is small for single
electrons at non-relativistic speed, it is drastically amplified
for certain superpositions of many-particle states.

Finally, we draw our conclusions in section 7.

\section{Reduced Density Matrix of the Matter Degrees of Freedom}
Our aim is to eliminate the variables of the electromagnetic
radiation field to obtain an exact representation for the reduced
density matrix $\rho_m$ of the matter degrees of freedom. The
starting point will be the following formal equation which relates
the density matrix $\rho_{m}(t_f)$ of the matter at some final
time $t_f$ to the density matrix $\rho(t_i)$ of the combined
matter-field system at some initial time $t_i$,
\begin{equation}
  \label{eq:1}
\rho_m(t_f) = \mathrm{tr}_{f}
 \left\{  T_{\leftarrow}
            \exp \left[ \int_{t_i}^{t_f} \mathrm{d}^4 x {\mathcal{L}}(x)
         \right] \rho(t_i)
     \right\}.
\end{equation}
The Liouville super-operator ${\mathcal{L}}(x)$ is defined as
\begin{equation}
  \label{eq:2}
  {\mathcal{L}}(x)\rho \equiv -i [{\mathcal{H}}(x), \rho ],
\end{equation}
where ${\mathcal{H}}(x)$ denotes the Hamiltonian density.
Space-time coordinates are written as $x^{\mu} = (x^0,\vec{x}) =
(t,\vec{x})$, where the speed of light $c$ is set equal to 1. All
fields are taken to be in the interaction picture and
$T_{\leftarrow}$ indicates the chronological time-ordering of the
interaction picture fields, while $\mathrm{tr}_f$ denotes the
trace over the variables of the radiation field. Setting
$\hbar=c=1$ we shall use here Heaviside-Lorentz units such that
the fine structure constant is given by
\begin{equation}
  \label{eq:3}
  \alpha = \frac{e^2}{ 4 \pi \hbar c} \approx \frac{1}{137}.
\end{equation}
To be specific we choose the Coulomb gauge in the following which
means that the Hamiltonian density takes the form
\cite{WEINBERG,JAUCH,COHEN}
\begin{equation}
  \label{eq:4}
  {\mathcal{H}}(x) =  {\mathcal{H}}_{\mathrm{C}}(x) +
  {\mathcal{H}}_{\mathrm{tr}}(x).
\end{equation}
Here,
\begin{equation}
 {\mathcal{H}}_{\mathrm{tr}}(x) =  j^{\mu}(x)A_{\mu}(x)
\end{equation}
represents the density of the interaction of the matter current
density $j^{\mu}(x)$ with the transversal radiation field,
\begin{equation}
  \label{eq:5}
  A^{\mu}(x) = (0, \vec{A}(x)), \qquad \vec{\nabla} \cdot \vec{A}(x)
  = 0,
\end{equation}
and
\begin{equation}
  \label{eq:6}
  {\mathcal{H}}_{\mathrm{C}}(x) = \frac{1}{2} j^0(x) A^0(x) = 
\frac{1}{2} \int \mathrm{d}^3y \;
  \frac{j^0(x^0,\vec{x})j^0(x^0,\vec{y})}{4 \pi |\vec{x} - \vec{y}|}
\end{equation}
is the Coulomb energy density such that
\begin{equation}
  \label{eq:7}
  H_C(x^0) = \frac{1}{2} \int \mathrm{d}^3x \int \mathrm{d}^3y
     \frac{j^0(x^0,\vec{x})j^0(x^0,\vec{y})}{4 \pi |\vec{x} - \vec{y}|}
\end{equation}
is the instantaneous Coulomb energy. Note that we use here the
convention that the electron charge $e$ is included in the current
density $j^{\mu}(x)$ of the matter.

Our first step is a decomposition of chronological time-ordering
operator $T_{\leftarrow}$ into a time-ordering operator
$T^j_{\leftarrow}$ for the matter current and a time-ordering
operator $T^A_{\leftarrow}$ for the electromagnetic field,
\begin{equation}
  \label{eq:8}
  T_{\leftarrow} = T^j_{\leftarrow} T^A_{\leftarrow}.
\end{equation}
This enables one to write Eq.~(\ref{eq:1}) as
\begin{equation}
  \label{eq:9}
  \rho_m(t_f) = T^j_{\leftarrow}
       \left( \mathrm{tr}_{f}
         \left\{  T^A_{\leftarrow} \exp
            \left[ \int_{t_i}^{t_f} \mathrm{d}^4x
              \left( {\mathcal{L}}_{\mathrm{C}}(x) +{\mathcal{L}}_{\mathrm{tr}}(x)  \right)
            \right] \rho(t_i)
          \right\}
       \right),
\end{equation}
where we have introduced the Liouville super-operators for the
densities of the Coulomb field and of the transversal field,
\begin{equation}
  \label{eq:10}
  {\mathcal{L}}_{\mathrm{C}}(x) \rho \equiv -\mathrm{i} [{\mathcal{H}}_{\mathrm{C}}(x), \rho ], \qquad
  {\mathcal{L}}_{\mathrm{tr}}(x) \rho \equiv -\mathrm{i} [j^{\mu}(x) A_{\mu}(x), \rho ].
\end{equation}
The currents $j^{\mu}$ commute under the time-ordering
$T^j_{\leftarrow}$. We may therefore treat them formally as
commuting $c$-number fields under the time-ordering symbol. Since
the super-operator ${\mathcal{L}}_{\mathrm{C}}(x)$ only contains
matter variables, the corresponding contribution can be pulled out
of the trace. Hence, we have
\begin{equation}
  \label{eq:11}
 \rho_m(t_f) =  T^j_{\leftarrow}
       \left( \exp
         \left[  \int_{t_i}^{t_f} \mathrm{d}^4x {\mathcal{L}}_{\mathrm{C}}(x) \right]
              \mathrm{tr}_{f}
           \left\{ T^A_{\leftarrow}
            \exp \left[  \int_{t_i}^{t_f} \mathrm{d}^4x {\mathcal{L}}_{\mathrm{tr}}(x)
                 \right] \rho(t_i)
           \right\}
       \right).
\end{equation}

We now proceed by eliminating the time-ordering of the $A$-fields.
With the help of the Wick-theorem we get
\begin{eqnarray}
  \label{eq:12}
\lefteqn{  T^A_{\leftarrow}
            \exp \left[  \int_{t_i}^{t_f} \mathrm{d}^4x {\mathcal{L}}_{\mathrm{tr}}(x)
                 \right] = } \\
& &  \exp \left[ \frac{1}{2} \int_{t_i}^{t_f} \mathrm{d}^4x
 \int_{t_i}^{t_f} \mathrm{d}^4x'[ {\mathcal{L}}_{\mathrm{tr}}(x), {\mathcal{L}}_{\mathrm{tr}}(x')]
 \theta(t-t') \right]
 \exp \left[   \int_{t_i}^{t_f} \mathrm{d}^4x  {\mathcal{L}}_{\mathrm{tr}}(x)
 \right]. \nonumber
\end{eqnarray}
In order to determine the commutator of the Liouville
super-operators we invoke the Jacobi identity which yields for an
arbitrary test density $\rho$,
\begin{eqnarray}
  \label{eq:13}
 [ {\mathcal{L}}_{\mathrm{tr}}(x), {\mathcal{L}}_{\mathrm{tr}}(x')] \rho &=&
       {\mathcal{L}}_{\mathrm{tr}}(x)   {\mathcal{L}}_{\mathrm{tr}}(x') \rho -
        {\mathcal{L}}_{\mathrm{tr}}(x')    {\mathcal{L}}_{\mathrm{tr}}(x) \rho  \nonumber \\
& = & - [ {\mathcal{H}}_{\mathrm{tr}}(x), [
{\mathcal{H}}_{\mathrm{tr}}(x'), \rho ]]
      + [{\mathcal{H}}_{\mathrm{tr}}(x')  , [ {\mathcal{H}}_{\mathrm{tr}}(x) ,\rho ]]  \nonumber \\
& = & - [ [ {\mathcal{H}}_{\mathrm{tr}}(x)
,{\mathcal{H}}_{\mathrm{tr}}(x') ],\rho].
\end{eqnarray}
The commutator of the transversal energy densities may be
simplified to read
\begin{equation}
  \label{eq:14}
   [ {\mathcal{H}}_{\mathrm{tr}}(x)  ,{\mathcal{H}}_{\mathrm{tr}}(x') ] =
     j^{\mu}(x) j^ {\nu}(x') [A_{\mu}(x), A_{\nu}(x')],
\end{equation}
since the contribution involving the commutator of the currents
vanishes by virtue of the time-ordering operator
$T^j_{\leftarrow}$. Thus, it follows from Eqs.~(\ref{eq:13}) and
(\ref{eq:14}) that the commutator of the Liouville super-operators
may be written as
\begin{equation}
  \label{eq:15}
 [ {\mathcal{L}}_{\mathrm{tr}}(x), {\mathcal{L}}_{\mathrm{tr}}(x')] \rho =
- [A_{\mu}(x), A_{\nu}(x')] [j^{\mu}(x) j^{\nu}(x'), \rho].
\end{equation}
It is useful to introduce current super-operators $J_+(x)$ and
$J_-(x)$ by means of
\begin{equation}
  \label{eq:16}
  J_+^{\mu}(x)\rho \equiv j^{\mu}(x) \rho, \qquad
  J_-^{\mu}(x) \rho \equiv \rho j^{\mu}(x).
\end{equation}
Thus, $J_+(x)$ is defined to be the current density acting from
the left, while $J_-(x)$ acts from the right on an arbitrary
density. With the help of these definitions we may write the
commutator of the Liouville super-operators as
\begin{eqnarray*}
  [ {\mathcal{L}}_{\mathrm{tr}}(x), {\mathcal{L}}_{\mathrm{tr}}(x')] =
&-& [A_{\mu}(x), A_{\nu}(x')] J_+^{\mu}(x) J_+^{\nu}(x') \\
&+&  [A_{\mu}(x), A_{\nu}(x')] J_-^{\mu}(x) J_-^{\nu}(x').
\end{eqnarray*}
Inserting this result into Eq.~(\ref{eq:12}), we can write
Eq.~(\ref{eq:11}) as
\begin{eqnarray}
  \label{eq:18}
  \rho_m(t_f) & = & T_{\leftarrow}^j
                   \left( \exp
                   \left[ \int_{t_i}^{t_f} \mathrm{d}^4 x {\mathcal{L}}_{\mathrm{C}}(x)
                         \right.  \right. \nonumber \\
 && \qquad  - \frac{1}{2}  \int_{t_i}^{t_f} \mathrm{d}^4 x 
           \int_{t_i}^{t_f} \mathrm{d}^4 x'
                        \theta(t-t') [A_{\mu}(x),A_{\nu}(x')]
                              J_+^{\mu}(x) J_+^{\nu}(x') \nonumber \\
 && \qquad  \left. + \frac{1}{2}  \int_{t_i}^{t_f} \mathrm{d}^4 x 
        \int_{t_i}^{t_f} \mathrm{d}^4 x'
                        \theta(t-t') [A_{\mu}(x),A_{\nu}(x')]
                              J_-^{\mu}(x) J_-^{\nu}(x') \right] \nonumber \\
 && \qquad \cdot {\mathrm{tr}}_f
    \left . \left\{ \exp \left[ \int_{t_i}^{t_f} \mathrm{d}^4 x
    {\mathcal{L}}_{\mathrm{tr}}(x) \right] \rho(t_i)  \right\} \right).
\end{eqnarray}
This is an exact formal representation for the reduced density
matrix of the matter variables. Note that the time-ordering of the
radiation degrees of freedom has been removed and that they enter
Eq.~(\ref{eq:18}) only through the functional
\begin{equation}
  \label{eq:19}
  W[J_+,J_-] \equiv {\mathrm{tr}}_f
             \left\{  \exp \left[ \int_{t_i}^{t_f} \mathrm{d}^4 x
                     {\mathcal{L}}_{\mathrm{tr}}(x) \rho(t_i) \right]
              \right\},
\end{equation}
since the commutator of the $A$-fields is a $c$-number function.

\section{The Influence Phase Functional of QED}
The functional (\ref{eq:19}) involves an average over the field
variables with respect to the initial state $\rho(t_i)$ of the
combined matter-field system. It therefore contains all
correlations in the initial state of the total system. Here, we
are interested in the destruction of coherence. Our central goal
is thus to investigate how correlations are built up through the
interaction between matter and radiation field. We therefore
consider now an initial state of low entropy which is given by a
product state of the form
\begin{equation}
  \label{eq:20}
  \rho(t_i) = \rho_m(t_i) \otimes \rho_f,
\end{equation}
where $\rho_m(t_i)$ is the density matrix of the matter at the
initial time and the density matrix of the radiation field
describes an equilibrium state at temperature $T$,
\begin{equation}
  \label{eq:21}
  \rho_f = \frac{1}{Z_f} \exp(- \beta H_f).
\end{equation}
Here, $H_f$ denotes the Hamiltonian of the free radiation field
and the quantity $Z_f = \mathrm{tr}_f [\exp( -\beta H_f)]$ is the
partition function with $\beta=1/k_BT$. In the following we shall
denote by
\begin{equation}
\langle \mathcal{O} \rangle_f \equiv \mathrm{tr}_f \left\{
\mathcal{O} \rho_f \right\}
\end{equation}
the average of some quantity $\mathcal{O}$ with respect to the
thermal equilibrium state (\ref{eq:21}).

The influence of the special choice (\ref{eq:20}) for the initial
condition can be eliminated by pushing $t_i \rightarrow - \infty$
and by switching on the interaction adiabatically. This is the
usual procedure used in Quantum Field Theory in order to define
asymptotic states and the $S$-matrix. The matter and the field
variables are then described as {\em{in}}-fields, obeying free
field equations with renormalized mass. These fields generate
physical one-particle states from the interacting ground state.

For an arbitrary initial condition $\rho(t_i)$ the functional
$W[J_+,J_-]$ can be determined, for example, by means of a
cumulant expansion. Since the initial state (\ref{eq:20}) is
Gaussian with respect to the field variables and since the
Liouville super-operator ${\mathcal{L}}_{\mathrm{tr}}(x)$ is
linear in the radiation field, the cumulant expansion terminates
after the second order term. In addition, a linear term does not
appear in the expansion because of $\langle A_{\mu}(x) \rangle_f
=0$. Thus we immediately obtain
\begin{equation}
  \label{eq:22}
  W[J_+,J_-] = \exp \left[ \frac{1}{2} \int_{t_i}^{t_f} \mathrm{d}^4x
                      \int_{t_i}^{t_f} \mathrm{d}^4x'
  \langle  {\mathcal{L}}_{\mathrm{tr}}(x){\mathcal{L}}_{\mathrm{tr}}(x') \rangle_f
                  \right] \rho_m(t_i).
\end{equation}
Inserting the definition for the Liouville super-operator
${\mathcal{L}}_{\mathrm{tr}}(x)$ into the exponent of this
expression one finds after some algebra,
\begin{eqnarray*}
  \lefteqn{ \frac{1}{2} \int_{t_i}^{t_f} \mathrm{d}^4x
                      \int_{t_i}^{t_f} \mathrm{d}^4x'
  \langle  {\mathcal{L}}_{\mathrm{tr}}(x){\mathcal{L}}_{\mathrm{tr}}(x') \rangle_f
                  \rho_m }   \\
&&\equiv - \frac{1}{2} \int_{t_i}^{t_f} \mathrm{d}^4x
                      \int_{t_i}^{t_f} \mathrm{d}^4x'  {\mathrm{tr}}_f
             \left\{
 [{\mathcal{H}}_{\mathrm{tr}}(x),[{\mathcal{H}}_{\mathrm{tr}}(x'),\rho_m \otimes \rho_f]]
             \right\}  \\
&& =- \frac{1}{2} \int_{t_i}^{t_f} \mathrm{d}^4x
                      \int_{t_i}^{t_f} \mathrm{d}^4x'
       \left[
          \langle A_{\nu}(x') A_{\mu}(x)\rangle_f J_+^{\mu}(x)
          J_+^{\nu}(x') \right. \\
&& \qquad \qquad \qquad \qquad \;\;\;\;\;
   + \langle A_{\mu}(x) A_{\nu}(x')\rangle_f J_-^{\mu}(x) J_-^{\nu}(x') \\
&& \qquad \qquad \qquad \qquad \;\;\;\;\;
   - \langle A_{\nu}(x') A_{\mu}(x)\rangle_f J_+^{\mu}(x) J_-^{\nu}(x') \\
&& \qquad \qquad \qquad \qquad \;\;\;\;\;
  \left. - \langle A_{\mu}(x) A_{\nu}(x')\rangle_f J_-^{\mu}(x) J_-^{\nu}(x')
     \right] \rho_m.
\end{eqnarray*}
On using this result together with Eq.~(\ref{eq:22}),
Eq.~(\ref{eq:18}) can be cast into the form,
\begin{eqnarray}
  \label{eq:24}
\lefteqn{\rho_m(t_f) = T_{\leftarrow}^j
          \Big( \exp
            \Big[ \int_{t_i}^{t_f} \mathrm{d}^4x {\mathcal{L}}_{\mathrm{C}}(x)
        } \\
&& \qquad \qquad + \frac{1}{2} \int_{t_i}^{t_f} \mathrm{d}^4x
                      \int_{t_i}^{t_f}  \mathrm{d}^4x'
          \big\{ - \big( \theta(t-t')[A_{\mu}(x), A_{\nu}(x')]
   \nonumber \\
&& \qquad \qquad \qquad \qquad \qquad \qquad \qquad \qquad
   + \langle A_{\nu}(x')A_{\mu}(x) \rangle_f
   \big) J_+^{\mu}(x) J_+^{\nu}(x')
   \nonumber \\
&& \qquad \qquad \qquad \qquad \qquad \qquad \;\;\;\;
   + \big( \theta(t-t')[A_{\mu}(x),A_{\nu}(x')]
   \nonumber \\
&& \qquad \qquad \qquad \qquad \qquad \qquad \qquad \qquad
   - \langle A_{\mu}(x)A_{\nu}(x') \rangle_f
   \big) J_-^{\mu}(x) J_-^{\nu}(x')
   \nonumber \\
&& \qquad \qquad \qquad \qquad \qquad \qquad \;\;\;\;
   + \langle A_{\nu}(x')A_{\mu}(x) \rangle_f   J_+^{\mu}(x) J_-^{\nu}(x')
   \nonumber \\
&& \qquad \qquad \qquad \qquad \qquad \qquad \;\;\;\;
   + \langle A_{\mu}(x)A_{\nu}(x') \rangle_f J_-^{\mu}(x) J_+^{\nu}(x')
   \big\} \Big] \Big) \rho_m(t_i). \nonumber
\end{eqnarray}

At this stage it is useful to introduce a new notation for the
correlation functions of the electromagnetic field, namely the
Feynman propagator and its complex conjugated ($T_{\rightarrow}$
denotes the anti-chronological time-ordering),
\begin{eqnarray}
  \label{eq:25}
  \mathrm{i} D_F(x-x')_{\mu \nu} &\equiv&
  \langle T_{\leftarrow} (A_{\mu}(x) A_{\nu}(x')) \rangle_f \nonumber \\
 &=& \theta(t-t')[A_{\mu}(x), A_{\nu}(x')] +
     \langle A_{\nu}(x')A_{\mu}(x) \rangle_f, \nonumber \\
  \mathrm{i} D_F^{\ast}(x-x')_{\mu \nu} &\equiv&
  -\langle T_{\rightarrow} (A_{\mu}(x) A_{\nu}(x')) \rangle_f \nonumber \\
 &=& \theta(t-t')[A_{\mu}(x), A_{\nu}(x')] -
     \langle A_{\mu}(x)A_{\nu}(x') \rangle_f,
\end{eqnarray}
as well as the two-point correlation functions
\begin{eqnarray}
  \label{eq:26}
  D_+(x-x')_{\mu \nu} &\equiv&  \langle
                  A_{\mu}(x) A_{\nu}(x') \rangle_f, \nonumber \\
  D_-(x-x')_{\mu \nu} &\equiv&  \langle
                  A_{\nu}(x') A_{\mu}(x) \rangle_f.
\end{eqnarray}
As is easily verified these functions are related through
\begin{equation}
  \label{eq:27}
  -\mathrm{i} D_F(x-x')_{\mu \nu} +\mathrm{i} D_F^{\ast}(x-x')_{\mu \nu}
  + D_+(x-x')_{\mu \nu} + D_-(x-x')_{\mu \nu} = 0.
\end{equation}
With the help of this notation the density matrix of the matter
can now be written as follows,
\begin{eqnarray}
  \label{eq:28}
 \lefteqn{\rho_m(t_f) = T_{\leftarrow}^j
          \left( \exp
            \left[ \int_{t_i}^{t_f} \mathrm{d}^4x {\mathcal{L}}_{c}(x)
              \right. \right.    } \\
&& \qquad \qquad \;\;\; \left. \left.
       +  \frac{1}{2} \int_{t_i}^{t_f} \mathrm{d}^4x
                      \int_{t_i}^{t_f} \mathrm{d}^4x'
          \left\{ -
           \mathrm{i} D_F(x-x')_{\mu \nu}
             J_+^{\mu}(x) J_+^{\nu}(x')
    \right. \right. \right. \nonumber  \\
&& \;\;\; \qquad \qquad \qquad \qquad \qquad \qquad \qquad
   + \mathrm{i} D_F^{\ast}(x-x')_{\mu \nu} J_-^{\mu}(x) J_-^{\nu}(x') \nonumber  \\
&& \;\;\; \qquad \qquad \qquad \qquad \qquad \qquad \qquad
   + D_-(x-x')_{\mu \nu} J_+^{\mu}(x) J_-^{\nu}(x') \nonumber \\
&& \;\;\; \qquad \qquad \qquad \qquad \qquad \qquad \qquad
   \left. \left. \left. \!\!
   + D_+(x-x')_{\mu \nu}   J_-^{\mu}(x) J_+^{\nu}(x')
   \right\} \right] \right) \rho_m(t_i). \nonumber
\end{eqnarray}
This equation provides an exact representation for the matter
density matrix which takes on the desired form: It involves the
electromagnetic field variables only through the various two-point
correlation functions introduced above. One observes that the
dynamics of the matter variables is given by a time-ordered
exponential function whose exponent is a bilinear functional of
the current super-operators $J_{\pm}(x)$. Formally we may write
Eq.~(\ref{eq:28}) as
\begin{equation}
  \label{eq:40}
  \rho_m(t_f) = T_{\leftarrow}^j \exp\left( \mathrm{i}\Phi[J_+,J_-] \right) \rho_m(t_i),
\end{equation}
where we have introduced an {\em{influence phase functional}}
\begin{eqnarray}
  \label{eq:42}
 \mathrm{i} \Phi[J_+,J_-] &=&  \int_{t_i}^{t_f}\mathrm{d}^4x {\mathcal{L}}_{\mathrm{C}}(x)
          + \frac{1}{2} \int_{t_i}^{t_f}\mathrm{d}^4x \int_{t_i}^{t_f}\mathrm{d}^4x' \\
  &&   \times \left\{
           -\mathrm{i} D_F(x-x')_{\mu \nu} J_+^{\mu}(x)   J_+^{\nu}(x')
          + \mathrm{i} D_F^{\ast}(x-x')_{\mu \nu} J_-^{\mu}(x)   J_-^{\nu}(x')
        \right. \nonumber \\
 && \;\;\;\;\; \left.
      + D_-(x-x')_{\mu \nu} J_+^{\mu}(x)   J_-^{\nu}(x')
          +D_+(x-x')_{\mu \nu} J_-^{\mu}(x)   J_+^{\nu}(x')
  \right\}. \nonumber
\end{eqnarray}
It should be remarked that the influence phase $\Phi[J_+,J_-]$ is
both a functional of the quantities $J_{\pm}(x)$ and a
super-operator which acts in the space of density matrices of the
matter degrees of freedom. There are several alternative methods
which could be used to arrive at an expression of the form
(\ref{eq:42}) as, for example, path integral techniques
\cite{FEYNMAN-VERNON} or Schwinger's closed time-path method
\cite{CHOU}. The expression (\ref{eq:42}) for the influence phase
functional has been given in Ref.~\cite{DIOSI} without the Coulomb
term and for the special case of zero temperature. In our
derivation we have combined super-operator techniques with methods
from field theory, which seems to be the most direct way to obtain
a representation of the reduced density matrix.

For the study of decoherence phenomena another equivalent formula
for the influence phase functional will be useful. To this end we
define the commutator function
\begin{eqnarray}
  \label{eq:29}
 D(x-x')_{\mu \nu} &\equiv& \mathrm{i} [A_{\mu}(x), A_{\nu}(x')] \nonumber \\
 &=& \mathrm{i} \left(D_+(x-x')_{\mu \nu} -D_-(x-x')_{\mu \nu}  \right)
\end{eqnarray}
and the anti-commutator function
\begin{eqnarray}
  \label{eq:30}
 D_1(x-x')_{\mu \nu} &\equiv&  \langle \{ A_{\mu}(x), A_{\nu}(x')\} \rangle_f
 \nonumber \\
 &=& D_+(x-x')_{\mu \nu} + D_-(x-x')_{\mu \nu}.
\end{eqnarray}
Of course, the previously introduced correlation functions may be
expressed in terms of $D(x-x')_{\mu \nu}$ and $D_1(x-x')_{\mu
\nu}$,
\begin{eqnarray}
  \label{eq:31}
D_+(x-x')_{\mu \nu} & = & \frac{1}{2} D_1(x-x')_{\mu \nu} -
                                \frac{\mathrm{i}}{2} D(x-x')_{\mu \nu}, \\
  \label{eq:32}
D_-(x-x')_{\mu \nu} & = & \frac{1}{2} D_1(x-x')_{\mu \nu}
                              +  \frac{\mathrm{i}}{2} D(x-x')_{\mu \nu}, \\
  \label{eq:33}
i D_F(x-x')_{\mu \nu} &=& \frac{1}{2} D_1(x-x')_{\mu \nu} -
            \frac{\mathrm{i}}{2} {\mathrm{sign}}(t-t') D(x-x')_{\mu \nu}, \\
 \label{eq:34}
- i D_F^{\ast}(x-x')_{\mu \nu} &=& \frac{1}{2} D_1(x-x')_{\mu \nu}
                 + \frac{\mathrm{i}}{2} {\mathrm{sign}}(t-t') D(x-x')_{\mu \nu}.
\end{eqnarray}
Correspondingly, we define a commutator super-operator $J_c(x)$
and an anti-commutator super-operator $J_a(x)$ by means of
\begin{equation}
  \label{eq:37}
  J_c^{\mu}(x) \rho \equiv [j^{\mu}(x), \rho],
\qquad
 J_a^{\mu}(x) \rho \equiv \{j^{\mu}(x), \rho\},
\end{equation}
which are related to the previously introduced super-operators
$J_{\pm}^{\mu}(x)$ by
\begin{equation}
  \label{eq:38}
   J_c^{\mu}(x)=  J_+^{\mu}(x) -  J_-^{\mu}(x),
\qquad
 J_a^{\mu}(x)=  J_+^{\mu}(x) +  J_-^{\mu}(x).
\end{equation}
In terms of these quantities the influence phase functional may
now be written as
\begin{eqnarray}
  \label{eq:39}
  \mathrm{i} \Phi[J_c,J_a] &=& \int_{t_i}^{t_f} 
        \mathrm{d}^4x {\mathcal{L}}_{\mathrm{C}}(x) \\
& &  + \int_{t_i}^{t_f} \! \mathrm{d}^4x \int_{t_i}^{t} \! \mathrm{d}^4x'
 \left\{ \frac{\mathrm{i}}{2} D(x-x')_{\mu \nu} J_c^{\mu}(x)  J_a^{\nu}(x')
   \right. \nonumber \\
& &  \qquad \qquad \qquad \qquad- \left.
       \frac{1}{2} D_1(x-x')_{\mu \nu} J_c^{\mu}(x) J_c^{\nu}(x')
         \right\}. \nonumber
\end{eqnarray}
This form of the influence phase functional will be particularly
useful later on. It represents the influence of the radiation
field on the matter dynamics in terms of the two fundamental
2-point correlation functions $D(x-x')$ and $D_1(x-x')$. Note that
the double space-time integral in Eq.~(\ref{eq:39}) is already a
time-ordered integral since the integration over $t'=x'^0$ extends
over the time interval from $t_i$ to $t=x^0$.

For a physical discussion of these results it may be instructive
to compare Eq.~(\ref{eq:28}) with the structure of a Markovian
quantum master equation in Lindblad form \cite{GARDINER},
\begin{equation}
\frac{d\rho_m}{dt} = -\mathrm{i}[H_m,\rho_m] + \sum_i \left( A_i \rho_m
A_i^{\dagger} -\frac{1}{2} A_i^{\dagger} A_i \rho_m - \frac{1}{2}
\rho_m A_i^{\dagger} A_i \right),
\end{equation}
where $H_m$ generates the coherent evolution and the $A_i$ denote
a set of operators, the Lindblad operators, labeled by some index
$i$. One observes that the terms of the influence phase functional
involving the current super-operators in the combinations $J_+J_-$
and $J_-J_+$ correspond to the gain terms in the Lindblad equation
having the form $A_i \rho_m A_i^{\dagger}$. These terms may be
interpreted as describing the back action on the reduced system of
the matter degrees of freedom induced by ``real'' processes in
which photons are absorbed or emitted. The presence of these terms
leads to a transformation of pure states into statistical
mixtures. Namely, if we disregard the terms containing the
combinations $J_+J_-$ and $J_-J_+$ the remaining expression takes
the form
\begin{equation}
  \label{eq:43}
  \rho_m(t_f) \approx U(t_f,t_i) \rho_m(t_i) U^{\dagger}(t_f,t_i),
\end{equation}
where
\begin{eqnarray}
  \label{eq:44}
 U(t_f,t_i) &=& T_{\leftarrow}^j \exp \left[
 -\mathrm{i} \int_{t_i}^{t_f} \mathrm{d}^4x {\mathcal{H}}_{\mathrm{C}}(x) \right. \\
 &~& \qquad \qquad \left. - \frac{\mathrm{i}}{2} \int_{t_i}^{t_f} \mathrm{d}^4x \int_{t_i}^{t_f} \mathrm{d}^4x'
                 D_F(x-x')_{\mu \nu} j^{\mu}(x) j^{\nu}(x')
                 \right]. \nonumber
\end{eqnarray}
Eq.~(\ref{eq:43}) shows that the contributions involving the
Feynman propagators and the combinations $J_+J_+$ and $J_-J_-$ of
super-operators preserve the purity of states \cite{DIOSI}. Recall
that all correlations functions have been defined in terms of the
transversal radiation field. We may turn to the covariant form of
the correlation functions if we replace at the same time the
current density by its transversal component
$j^{\mu}_{\mathrm{tr}}$. The expression (\ref{eq:44}) is then seen
to contain the vacuum-to-vacuum amplitude $A[j]$ of the
electromagnetic field in the presence of a classical, transversal
current density $j^{\mu}_{\mathrm{tr}}(x)$ \cite{FEYNMAN},
\begin{equation}
  \label{eq:45}
  A[j] = \exp\left[
  - \frac{\mathrm{i}}{2} \int \mathrm{d}^4x \int \mathrm{d}^4x'
  D_F(x-x')_{\mu \nu} j^{\mu}_{\mathrm{tr}}(x) j^{\nu}_{\mathrm{tr}}(x')
  \right].
\end{equation}
With the help of the decomposition (\ref{eq:33}) of the Feynman
propagator into a real and an imaginary part we find
\begin{equation}
  \label{eq:46}
  A[j] = \exp \left[\mathrm{i} \left( S^{(1)} + \mathrm{i} S^{(2)} \right) \right].
\end{equation}
The vacuum-to-vacuum amplitude is thus represented in terms of a
complex action functional with the real part
\begin{equation}
  \label{eq:47}
  S^{(1)} = \frac{1}{4} \int \mathrm{d}^4x \int \mathrm{d}^4x'
    {\mathrm{sign}}(t-t') D(x-x')_{\mu \nu} j^{\mu}_{\mathrm{tr}}(x)
    j^{\nu}_{\mathrm{tr}}(x'),
\end{equation}
and with the imaginary part
\begin{equation}
  \label{eq:48}
  S^{(2)} = \frac{1}{4}  \int \mathrm{d}^4x \int \mathrm{d}^4x'
  D_1(x-x')_{\mu \nu} j^{\mu}_{\mathrm{tr}}(x) j^{\nu}_{\mathrm{tr}}(x').
\end{equation}
The imaginary part $S^{(2)}$ yields the probability that no photon
is emitted by the current $j^{\mu}_{\mathrm{tr}}$,
\begin{equation}
  \label{eq:49}
  | A[j]|^2 = \exp \left(- 2 S^{(2)}\right).
\end{equation}
In covariant form we have
\begin{equation}
  \label{eq:50}
  D(x-x')_{\mu \nu} = - \frac{1}{2 \pi}
              {\mathrm{sign}}(t-t') \delta[(x-x')^2] g_{\mu \nu},
\end{equation}
and, hence,
\begin{equation}
\label{eq:51} S^{(1)} = - \frac{1}{8 \pi} \int \mathrm{d}^4x \int \mathrm{d}^4x'
\delta[(x-x')^2] j_{\mu}^{\mathrm{tr}}(x)
j^{\mu}_{\mathrm{tr}}(x').
\end{equation}
This is the classical Feynman-Wheeler action. It describes the
classical motion of a system of charged particles by means of a
non-local action which arises after the elimination of the degrees
of freedom of the electromagnetic radiation field. In the
following sections we will demonstrate that it is just the
imaginary part $S^{(2)}$ which leads to the destruction of
coherence of the matter degrees of freedom.

\section{The Interaction of a Single Electron with the Radiation Field}
In this section we shall apply the foregoing general theory to the
case of a single electron interacting with the radiation field
where we confine ourselves to the non-relativistic approximation.
It will be seen that this simple case already contains the basic
physical mechanism leading to decoherence.

\subsection{Representation of the Electron Density Matrix in the
            Non-Relativistic Approximation}
The starting point will be the representation (\ref{eq:40}) for
the reduced matter density with expression (\ref{eq:39}) for the
influence phase functional $\Phi$. It must be remembered that the
correlation functions $D(x-x')_{\mu \nu}$ and $D_1(x-x')_{\mu
\nu}$ have been defined in terms of the transversal radiation
field using Coulomb gauge and that they thus involve projections
onto the transversal component. In fact, we have the replacements,
\[
 D(x-x')_{\mu \nu} \longrightarrow D(x-x')_{ij} =
               - \left( \delta_{ij} - \frac{\partial_i \partial_j}{\Delta}
                 \right)  D(x-x')
\]
for the commutator functions, and
\[
 D_1(x-x')_{\mu \nu} \longrightarrow  D_1(x-x')_{ij} =
 + \left( \delta_{ij} - \frac{\partial_i \partial_j}{\Delta}
                 \right)  D_1(x-x')
\]
for the anti-commutator function, where
\[
 D(x-x') = -\mathrm{i} \int \frac{\mathrm{d}^3k}{2(2\pi)^3\omega}
         \left[ \exp\left(-\mathrm{i}k(x-x')\right) -  
            \exp\left(\mathrm{i}k(x-x')\right)
         \right],
\]
and
\[
 D_1(x-x') = \int \frac{\mathrm{d}^3k}{2(2\pi)^3\omega}
         \left[ \exp\left(-\mathrm{i}k(x-x')\right) +  
          \exp\left(\mathrm{i}k(x-x')\right)
         \right] \coth \left( \beta\omega/2 \right),
\]
with the notation $k^{\mu}=(\omega,\vec{k})=(|\vec{k}|,\vec{k})$
for the components of the wave vector. It should be noted that the
commutator function is independent of the temperature, while the
anti-commutator function does depend on $T$ through the factor
$\coth(\beta \omega/2) = 1 + 2 N(\omega)$, where $N(\omega)$ is
the average number of photons in a mode with frequency $\omega$.
Hence, invoking the non-relativistic (dipole) approximation we may
replace
\begin{equation}
  \label{eq:56}
  D(x-x')_{ij} \longrightarrow D(t-t')_{ij} = \delta_{ij} D(t-t')
  = \delta_{ij} \int_0^{\infty} \mathrm{d} \omega J(\omega) \sin \omega(t-t'),
\end{equation}
and
\begin{eqnarray}
  \label{eq:57}
  D_1(x-x')_{ij} \longrightarrow D_1(t-t')_{ij} &=& \delta_{ij}
  D_1(t-t') \\
  &=& \delta_{ij} \int_0^{\infty} \mathrm{d} \omega J(\omega)
      \coth \left( \beta\omega/2 \right) \cos \omega(t-t'),
      \nonumber
\end{eqnarray}
where we have introduced the spectral density
\begin{equation}
  \label{eq:58}
  J(\omega) = \frac{e^2}{3 \pi^2} \omega \Theta(\Omega - \omega),
\end{equation}
with some ultraviolet cutoff $\Omega$ (see below). It is important
to stress here that the spectral density increases with the first
power of the frequency $\omega$. Had we used dipole coupling $-e
\vec{x} \cdot \vec{E}$ of the electron coordinate $\vec{x}$ to the
electric field strength $\vec{E}$, the corresponding spectral
density would be proportional to the third power of the frequency.
This means that the coupling to the radiation field in the dipole
approximation may be described as a special case of the famous
Caldeira-Leggett model \cite{LEGGETT} and that in the language of
the theory of quantum Brownian motion \cite{GRABERT} the radiation
field constitutes a super-Ohmic environment
\cite{CALDEIRA,ZUREK97}. Note also that we now include the factor
$e^2$ into the definition of the correlation function. Within the
non-relativistic approximation we may thus replace the current
density by
\begin{equation}
  \label{eq:59}
  \vec{j}(t,\vec{x}) \longrightarrow \frac{\vec{p}(t)}{2m}
  \delta(\vec{x}-\vec{x}(t)) +
  \delta(\vec{x}-\vec{x}(t)) \frac{\vec{p}(t)}{2m},
\end{equation}
where $\vec{p}(t)$ and $\vec{x}(t)$ denote the momentum and
position operator of the electron in the interaction picture with
respect to the Hamiltonian
\begin{equation}
  \label{eq:61}
  H_m = \frac{\vec{p}^2}{2m} + V(\vec{x})
\end{equation}
for the electron, $V(\vec{x})$ being some external potential.

We are thus led to the following non-relativistic approximation of
Eq.~(\ref{eq:40}),
\begin{eqnarray}
  \label{eq:60}
 \rho_m(t_f) & = & T_{\leftarrow}
      \left( \exp \left[ \int_{t_i}^{t_f} \mathrm{d}t
                           \int_{t_i}^{t} \mathrm{d}t'
                  \left\{ \frac{\mathrm{i}}{2} D(t-t')
                           \frac{\vec{p}_c(t)}{m}  \frac{\vec{p}_a(t')}{m}
       \right. \right. \right.  \\
  & & \qquad \qquad \qquad \qquad \;\;\;\;\;\;\;
       \left. \left. \left. - \frac{1}{2} D_1(t-t')
                  \frac{\vec{p}_c(t)}{m}  \frac{\vec{p}_c(t')}{m}
         \right\} \right] \right) \rho_m(t_i). \nonumber
\end{eqnarray}
This equation represents the density matrix (neglecting the spin
degree of freedom) for a single electron interacting with the
radiation field at temperature $T$. In accordance with the
definitions (\ref{eq:37}) and (\ref{eq:38}) $\vec{p}_c$ is a
commutator super-operator and $\vec{p}_a$ an anti-commutator
super-operator. In the theory of quantum Brownian motion the
function $D(t-t')$ is called the {\em{dissipation kernel}},
whereas $D_1(t-t')$ is referred to as {\em{noise kernel}}.

\subsection{The Path Integral Representation}
The reduced density matrix given in Eq.~(\ref{eq:60}) admits an
equivalent path integral representation \cite{GRABERT} which may
be written as follows,
\begin{equation}
  \label{eq:62}
  \rho_m(\vec{x}_f, \vec{x}'_f,t_f) =
       \int \mathrm{d}^3x_i \int \mathrm{d}^3 x_i' J(\vec{x}_f, \vec{x}'_f,t_f;
                                   \vec{x}_i, \vec{x}'_i,t_i)
                    \rho_m(\vec{x}_i, \vec{x}'_i,t_i),
\end{equation}
with the {\em{propagator function}}
\begin{equation}
  \label{eq:63}
 J(\vec{x}_f, \vec{x}'_f,t_f; \vec{x}_i, \vec{x}'_i,t_i) =
\int \mathrm{D} \vec{x} \mathrm{D} \vec{x}' \exp
   \left\{  \mathrm{i} \left( S_m[\vec{x}] - S_m[\vec{x}']  \right)
            + \mathrm{i} \Phi[\vec{x},\vec{x}'] \right\}.
\end{equation}
This is a double path integral which is to be extended over all
paths $\vec{x}(t)$ and $\vec{x}'(t)$ with the boundary conditions
\begin{equation}
  \label{eq:64}
\vec{x}'(t_i) = \vec{x}'_i, \qquad  \vec{x}'(t_f) = \vec{x}'_f,
\qquad   \vec{x}(t_i) = \vec{x}_i, \qquad   \vec{x}(t_f) =
\vec{x}_f.
\end{equation}
$S_{m}[\vec{x}]$ denotes the action functional for the electron,
\begin{equation}
  \label{eq:65}
  S_m[\vec{x}] = \int_{t_i}^{t_f} \mathrm{d}t
    \left(\frac{1}{2} m \vec{\dot{x}}^2 - V(\vec{x}) \right),
\end{equation}
while the influence phase functional becomes,
\begin{eqnarray}
  \label{eq:66}
  \mathrm{i} \Phi[\vec{x}, \vec{x}'] & = &
 \int_{t_i}^{t_f} \mathrm{d}t  \int_{t_i}^{t} \mathrm{d}t'
      \left\{ \frac{\mathrm{i}}{2} D(t-t')
              \left( \dot{\vec{x}}(t) - \dot{\vec{x}}'(t) \right)
             \left( \dot{\vec{x}}(t') + \dot{\vec{x}}'(t') \right)
       \right.
       \nonumber \\
& & \left. \qquad \qquad \;\;
   - \frac{1}{2} D_1(t-t')
            \left( \dot{\vec{x}}(t) - \dot{\vec{x}}'(t) \right)
             \left( \dot{\vec{x}}(t') - \dot{\vec{x}}'(t') \right)
    \right\}.
\end{eqnarray}
We define the new variables
\begin{equation}
  \label{eq:67}
  \vec{q} = \vec{x} - \vec{x}', \qquad
\vec{r} = \frac{1}{2} (\vec{x} + \vec{x}'),
\end{equation}
and set, for simplicity, the initial time equal to zero, $t_i=0$.
We may then write Eq. (\ref{eq:62}) as
\begin{equation}
  \label{eq:69}
  \rho_m(\vec{r}_f, \vec{q}_f,t_f) =
\int \mathrm{d}^3r_i \int \mathrm{d}^3q_i
 J(\vec{r}_f, \vec{q}_f, t_f;\vec{r}_i, \vec{q}_i)
     \rho_m(\vec{r}_i, \vec{q}_i,0).
\end{equation}
The propagator function
\begin{equation}
  \label{eq:70}
 J(\vec{r}_f, \vec{q}_f, t_f;\vec{r}_i, \vec{q}_i) =
 \int \mathrm{D} \vec{r} \int \mathrm{D}\vec{q} \exp \{ \mathrm{i}
{\mathcal{A}}[\vec{r},\vec{q}]  \}
\end{equation}
is a double path integral over all path $\vec{r}(t)$, $\vec{q}(t)$
satisfying the boundary conditions,
\begin{equation}
 \label{eq:115}
 \vec{r}(0) = \vec{r}_i, \qquad
 \vec{r}(t_f) = \vec{r}_f, \qquad
 \vec{q}(0) = \vec{q}_i, \qquad
 \vec{q}(t_f) = \vec{q}_f.
\end{equation}
The weight factor for the paths $\vec{r}(t)$, $\vec{q}(t)$ is
defined in terms of an effective action ${\mathcal{A}}$
functional,
\begin{eqnarray}
  \label{eq:71}
 {\mathcal{A}}[\vec{r},\vec{q}]& =& \int_0^{t_f} \mathrm{d}t
 \left( m \dot{\vec{r}} \dot{\vec{q}} - V(\vec{r} + \frac{1}{2} \vec{q})
                              + V(\vec{r} - \frac{1}{2} \vec{q})   \right)
                    \nonumber \\
  && + \int_0^{t_f} \mathrm{d}t \int_0^{t_f} \mathrm{d}t'
          \theta(t-t') D(t-t') \dot{\vec{q}}(t) \dot{\vec{r}}(t')
                   \nonumber \\
  && + \frac{\mathrm{i}}{4}    \int_0^{t_f} \mathrm{d}t \int_0^{t_f} \mathrm{d}t'
                  D_1(t-t')    \dot{\vec{q}}(t)   \dot{\vec{q}}(t').
\end{eqnarray}
The first variation of ${\mathcal{A}}$ is found to be
\begin{eqnarray}
  \label{eq:72}
  \delta {\mathcal{A}} &=& - \int_0^{t_f} \mathrm{d}t
    \left\{  \delta \vec{q}(t)
        \left[ m \ddot{\vec{r}}(t) +
                 \frac{1}{2} \vec{\nabla}_{\vec{r}}
                        (V(\vec{r} + \frac{1}{2} \vec{q})
                            + V(\vec{r} - \frac{1}{2} \vec{q})) \right.
    \right.
           \nonumber \\
&& \left. \left. \qquad \qquad \qquad
          + \frac{\mathrm{d}}{\mathrm{d}t} \int_0^t \mathrm{d}t' D(t-t') \dot{\vec{r}}(t')
         + \frac{\mathrm{i}}{2} \frac{\mathrm{d}}{\mathrm{d}t}
                   \int_0^{t_f} \mathrm{d}t' D_1(t-t') \dot{\vec{q}}(t')
    \right] \right.
         \nonumber \\
&&  \left. \qquad \qquad
     + \delta \vec{r}(t) \left[
             m \ddot{\vec{q}}(t) +
               2 \vec{\nabla}_{\vec{q}}
                        (V(\vec{r} + \frac{1}{2} \vec{q})
                      + V(\vec{r} - \frac{1}{2} \vec{q})) \right. \right. \nonumber \\
&& \left. \left. \qquad \qquad \qquad
                       + \frac{\mathrm{d}}{\mathrm{d}t}
                   \int_t^{t_f} \mathrm{d}t' D(t'-t) \dot{\vec{q}}(t')
  \right] \right\},
\end{eqnarray}
which leads to the classical equations of motion,
\begin{eqnarray}
  \label{eq:73}
 && m \ddot{\vec{r}}(t) +  \frac{1}{2} \vec{\nabla}_{\vec{r}}
                        (V(\vec{r} + \frac{1}{2} \vec{q})
                            + V(\vec{r} - \frac{1}{2} \vec{q}))
 + \frac{\mathrm{d}}{\mathrm{d}t} \int_0^t \mathrm{d}t' D(t-t') \dot{\vec{r}}(t') \nonumber \\
 && = - \frac{\mathrm{i}}{2}  \frac{\mathrm{d}}{\mathrm{d}t}
                   \int_0^{t_f} \mathrm{d}t' D_1(t-t') \dot{\vec{q}}(t'),
\end{eqnarray}
and
\begin{equation}
  \label{eq:74}
 m \ddot{\vec{q}}(t) +
               2 \vec{\nabla}_{\vec{q}}
                        (V(\vec{r} + \frac{1}{2} \vec{q})
                      + V(\vec{r} - \frac{1}{2} \vec{q}))
                       + \frac{\mathrm{d}}{\mathrm{d}t}
                   \int_t^{t_f} \mathrm{d}t' D(t'-t) \dot{\vec{q}}(t')
    =0.
\end{equation}

\subsection{The Abraham-Lorentz Equation}
The real part of the equation of motion (\ref{eq:73}), which is
obtained by setting the right-hand side equal to zero, yields the
famous Abraham-Lorentz equation for the electron \cite{JACKSON}.
It describes the radiation damping through the damping kernel
$D(t-t')$ \cite{CALDEIRA}. To see this we write the real part of
Eq.~(\ref{eq:73}) as
\begin{equation}
  \label{eq:75}
 m \ddot{\vec{r}}(t)   + \frac{\mathrm{d}}{\mathrm{d}t} \int_0^t 
\mathrm{d}t' D(t-t') \dot{\vec{r}}(t')
= \vec{F}_{\mathrm{ext}}(t),
\end{equation}
where $\vec{F}_{\mathrm{ext}}(t)$ denotes an external force
derived from the potential $V$. The damping kernel can be written
(see Eqs.~(\ref{eq:56}) and (\ref{eq:58}))
\begin{eqnarray*}
  D(t-t') &=& \int_0^{\Omega} \mathrm{d} \omega
                \frac{e^2}{3 \pi^2} \omega \sin \omega (t-t')
           =  \frac{e^2}{3 \pi^2} \frac{\mathrm{d}}{\mathrm{d}t'}
               \int_0^{\Omega} \mathrm{d} \omega  \cos \omega(t-t') \\
         & = &  \frac{e^2}{3 \pi^2} \frac{\mathrm{d}}{\mathrm{d}t'}
                        \frac{\sin \Omega (t-t')}{t-t'}
         \equiv \frac{e^2}{3 \pi^2}  \frac{\mathrm{d}}{\mathrm{d}t'} f(t-t'),
\end{eqnarray*}
where we have introduced the function
\begin{equation}
  \label{eq:77}
  f(t) \equiv \frac{\sin \Omega t}{t}.
\end{equation}
To be specific the UV-cutoff $\Omega$ is taken to be
\begin{equation}
  \label{eq:78}
  \hbar \Omega = m c^2,
\end{equation}
which implies that
\begin{equation}
  \label{eq:79}
  \Omega = \frac{m c^2}{\hbar} = \frac{c}{\bar{\lambda}_C},
\end{equation}
where 
\begin{equation}
  \label{eq:80}
  \bar{\lambda}_C = \frac{\hbar}{m c}
\end{equation}
is the Compton wavelength.
For an electron we have
\begin{equation}
  \label{eq:81}
  \bar{\lambda}_C \approx 3.8 \times 10^{-13} {\mathrm{m}} \qquad
\mathrm{and}  \qquad \Omega \approx 0.78 \times 10^{21}
{\mathrm{s}}^{-1}.
\end{equation}

The term of the equation of motion (\ref{eq:75}) involving the
damping kernel can be written as follows,
\begin{eqnarray}
\label{eq:83} \lefteqn{\frac{\mathrm{d}}{\mathrm{d}t} \int_0^t dt' D(t-t')
\dot{\vec{r}}(t')  =
    \frac{e^2}{3 \pi^2} \frac{\mathrm{d}}{\mathrm{d}t}
             \int_0^t \mathrm{d}t' \left[ \frac{\mathrm{d}}{\mathrm{d}t'} f(t-t')  \right]
             \dot{\vec{r}}(t')} \\
& = &  \frac{e^2}{3 \pi^2} \frac{\mathrm{d}}{\mathrm{d}t}
          \left[ -  \int_0^t \mathrm{d}t'f(t-t') \ddot{\vec{r}}(t')
                  + f(0) \dot{\vec{r}}(t) - f(t) \dot{\vec{r}}(0).
                    \right] \nonumber
\end{eqnarray}
For times $t$ such that $\Omega t \gg 1$, i.e. $t \gg 10^{-21}$s,
we may replace
\begin{equation}
  \label{eq:82}
  f(t) \longrightarrow \pi \delta(t),
\end{equation}
and approximate $f(t) \approx 0$, while Eq.~(\ref{eq:77}) yields
$f(0)= \Omega$. Thus we obtain,
\begin{eqnarray}
  \label{eq:84}
 \frac{\mathrm{d}}{\mathrm{d}t} \int_0^t \mathrm{d}t' D(t-t') \dot{\vec{r}}(t') & = &
         \frac{e^2}{3 \pi^2} \frac{\mathrm{d}}{\mathrm{d}t}
               \left[ - \frac{\pi}{2} \ddot{\vec{r}}(t)
                 + \Omega \dot{\vec{r}}(t) \right],
\end{eqnarray}
which finally leads to the equation of motion,
\begin{equation}
  \label{eq:85}
  \left( m + \frac{e^2 \Omega}{3 \pi^2} \right) \dot{\vec{v}}(t)
     - \frac{e^2}{6 \pi} \ddot{\vec{v}}(t) =
 \vec{F}_{\mathrm{ext}}(t),
\end{equation}
where $\vec{v}=\dot{\vec{r}}$ is the velocity. This is the famous
Abraham-Lorentz equation \cite{JACKSON}. The term proportional to
the third derivative of $\vec{r}(t)$ describes the damping of the
electron motion through the emitted radiation. This term does not
depend on the cutoff frequency, while the cutoff-dependent term
yields a renormalization of the electron mass,
\begin{equation}
  \label{eq:88}
 m_R = m + \Delta m = m + \frac{e^2 \Omega}{3 \pi^2}.
\end{equation}
It is important to note that the electro-magnetic mass $\Delta m$
diverges linearly with the cutoff. The equation of motion
(\ref{eq:85}) can be obtained heuristically by means of the Larmor
formula for the power radiated by an accelerated charge. More
rigorously, it has been derived by Abraham and by Lorentz from the
conservation law for the field momentum, assuming a spherically
symmetric charge distribution and that the momentum is of purely
electromagnetic origin \cite{JACKSON}.

For the cutoff $\Omega$ chosen above we get
\begin{equation}
  \label{eq:89}
  \Delta m = \frac{m e^2}{3 \pi^2} = \frac{4}{3 \pi} \alpha m,
\end{equation}
and, hence,
\begin{displaymath}
  \frac{\Delta m}{m} = \frac{4}{3 \pi} \alpha \approx 0.0031.
\end{displaymath}
The decomposition (\ref{eq:88}) of the mass is, however,
unphysical, since the electron is never observed without its
self-field and the associated field momentum. In other words, we
have to identify the renormalized mass $m_R$ with the observed
physical mass which enables us to write Eq. (\ref{eq:85}) as
\begin{equation}
  \label{eq:91}
  m_R \left[ \dot{\vec{v}}(t) - \tau_0 \ddot{\vec{v}}(t) \right]
           = \vec{F}_{\mathrm{ext}}(t).
\end{equation}
Here, the radiation damping term has been written in terms of a
characteristic radiation time scale $\tau_0$ given by
\begin{equation}
  \label{eq:92}
  \tau_0 \equiv \frac{e^2}{6 \pi m_R} = \frac{2}{3} r_e
  \approx 0.6 \times 10^{-23} \mathrm{s},
\end{equation}
where $r_e$ denotes the classical electron radius,
\begin{equation}
  \label{eq:93}
  r_e = \frac{e^2}{4 \pi m_R} = \alpha \bar{\lambda}_C
  \approx 2.8 \times 10^{-15} \mathrm{m}.
\end{equation}

It is well-known that Eq.~(\ref{eq:91}), being a classical
equation of motion for the electron, leads to the problem of
exponentially increasing {\em{runaway solutions}}. Namely, for
$\vec{F}_{\mathrm{ext}}=0$ we have
\begin{equation}
  \label{eq:95}
  \dot{\vec{v}} - \tau_0 \ddot{\vec{v}} = 0.
\end{equation}
In addition to the trivial solution of a constant velocity,
$\vec{v}={\mathrm{const}}$, one also finds the solution
\begin{displaymath}
  \dot{\vec{v}}(t) = \dot{\vec{v}}(0) \exp(t/\tau_0),
\end{displaymath}
describing an exponential growth of the acceleration for
$\dot{\vec{v}}(0) \neq 0$. In order to exclude these solutions one
imposes the boundary condition
\begin{displaymath}
  \dot{\vec{v}}(t) \longrightarrow 0 \qquad \mathrm{for} \qquad
  t \longrightarrow  \infty,
\end{displaymath}
if $\vec{F}_{\mathrm{ext}}$ also vanishes in this limit.  This
boundary condition can be implemented by rewriting
Eq.~(\ref{eq:91}) as an integro-differential equation
\begin{equation}
    \label{eq:96}
m_R \ddot{\vec{r}}(t) = \int_0^{\infty} \mathrm{d}s \exp(-s)
                   \vec{F}_{\mathrm{ext}}(t+\tau_0 s).
\end{equation}
On differentiating Eq.~(\ref{eq:96}) with respect to time, it is
easily verified that one is led back to Eq.~(\ref{eq:91}).
However, for $\vec{F}_{\mathrm{ext}}=0$ it follows immediately
from Eq.~(\ref{eq:96}) that $\vec{v} = \mathrm{const}$, such that
runaway solutions are excluded.

On the other hand, Eq.~(\ref{eq:96}) shows that the acceleration
depends upon the future value of the force. Hence, the electron
reacts to signals lying a time of order $\tau_0$ in the future,
which is the phenomenon of pre-acceleration. This phenomenon
should, however, not be taken too seriously, since the description
is only classical. The time scale $\tau_0$ corresponds to a length
scale $r_e$ which is smaller than the Compton wavelength
$\bar{\lambda}_C$ by a factor of $\alpha$, such that a quantum
mechanical treatment of the problem is required.

\subsection{Construction of the Decoherence Function}
In this subsection we derive the explicit form of the propagator
function (\ref{eq:70}) for the reduced electron density matrix in
the case of quadratic potentials,
\begin{equation}
  \label{eq:98}
  V(\vec{x}) = \frac{1}{2} m_R \omega_0^2 \vec{x}^2.
\end{equation}
Our aim is to introduce and to determine the {\em{decoherence
function}} which provides a quantitative measure for the degree of
decoherence. On using
\begin{eqnarray}
  \label{eq:99}
  V(\vec{r} + \vec{q}/2) + V(\vec{r} - \vec{q}/2)
   & =& m_R \omega_0^2 \vec{r}^2 + m_R \omega_0^2 \vec{q}^2/4 \\
 - V(\vec{r} + \vec{q}/2) + V(\vec{r} - \vec{q}/2)
   & = & -m_R \omega_0^2 \vec{r}\cdot \vec{q},
       \nonumber
\end{eqnarray}
the classical equations of motion take the form
\begin{eqnarray}
  \label{eq:100}
 m_R \left[\ddot{\vec{r}}(t)
           + \omega_0^2 \int_0^{\infty} \!\!\!\! \mathrm{d}s \exp(-s) \vec{r}(t+ \tau_0 s)
            \right] & = & \!\!
    - \frac{\mathrm{i}}{2} \frac{\mathrm{d}}{\mathrm{d}t} \int_0^{t_f} 
   \!\!\!\! 
\mathrm{d}t' D_1(t-t') \dot{\vec{q}}(t') \\
  \label{eq:101}
 m_R \left[\ddot{\vec{q}}(t)
           + \omega_0^2 \int_0^{\infty} \!\!\!\! \mathrm{d}s \exp(-s) \vec{q}(t- \tau_0 s)
            \right] & = & 0.
\end{eqnarray}
Note, that Eq.~(\ref{eq:101})  is the {\em{backward equation}} of
the real part of Eq.~(\ref{eq:100}). More precisely, if
$\vec{q}(t)$ solves Eq.~(\ref{eq:101}), then $\vec{r}(t) \equiv
\vec{q}(t_f -t)$ is a solution of (\ref{eq:100}) with the
right-hand side set equal to zero.

The above equations of motion lead to the following renormalized
action functional
\begin{eqnarray}
  \label{eq:102}
  {\mathcal{A}}[\vec{r}, \vec{q}] & = &
               \int_0^{t_f} \mathrm{d}t \, m_R
            \left[ \dot{\vec{r}}(t) \dot{\vec{q}}(t)
                - \omega_0^2 \vec{q}(t)
                  \int_0^{\infty} \mathrm{d}s \exp(-s) \vec{r}(t + \tau_0 s) \right]
                \nonumber \\
& & + \frac{\mathrm{i}}{4} \int_0^{t_f} \mathrm{d}t  \int_0^{t_f} \mathrm{d}t'
             D_1(t-t') \dot{\vec{q}}(t) \dot{\vec{q}}(t').
\end{eqnarray}
In the following we shall use this renormalised action functional
instead of the action given in Eq.~(\ref{eq:71}). By variation
with respect to $\vec{q}(t)$ we immediately obtain
Eq.~(\ref{eq:100}), whereas the variation with respect to
$\vec{r}(t)$ yields:
\begin{displaymath}
  - \int_0^{t_f} \mathrm{d}t \; m_R
   \left[ \ddot{\vec{q}}(t) \delta \vec{r}(t) +
        \omega_0^2 \int_0^{\infty} \mathrm{d}s \exp(-s)
         \vec{q}(t) \delta \vec{r}(t+\tau_0 s)  \right] =0,
\end{displaymath}
which implies
\begin{eqnarray}
  \label{eq:103}
\lefteqn{ \int_0^{t_f} \mathrm{d}t \ddot{\vec{q}}(t) \delta \vec{r}(t)
      + \omega_0^2 \int_0^{\infty} \mathrm{d}s \int_0^{t_f} \mathrm{d}t
         \exp(-s) \vec{q}(t) \delta \vec{r}(t+ \tau_0 s) } \nonumber \\
& = &\int_0^{t_f} \mathrm{d}t \ddot{\vec{q}}(t) \delta \vec{r}(t) +
    \omega_0^2 \int_0^{\infty} \mathrm{d}s \int_{\tau_0 s}^{t_f + \tau_0 s} 
 \mathrm{d}t
         \exp(-s) \vec{q}(t-\tau_0 s) \delta \vec{r}(t) \nonumber \\
& = & 0.
\end{eqnarray}
In the last time integral we may extend the integration over the
time interval from 0 to $t_f$. This is legitimate since $\tau_0$
is the radiation time scale: By setting this variation of the
action equal to zero we thus neglect times of the order of the
pre-acceleration time, which directly leads to the equation of
motion (\ref{eq:101}).

Since the action functional is quadratic the propagator function
can be determined exactly by evaluating the action along the
classical solution and by taking into account Gaussian
fluctuations around the classical paths. We therefore assume that
$\vec{r}(t)$ and $\vec{q}(t)$ are solutions of the classical
equations of motion (\ref{eq:100}) and (\ref{eq:101}) with
boundary conditions (\ref{eq:115}). The effective action along
these solutions may be written as
\begin{eqnarray}
  \label{eq:104}
{\mathcal{A}}_{\mathrm{cl}}[\vec{r}, \vec{q}] & = &
         m_R [ \dot{\vec{r}}_f \vec{q}_f
               -\dot{\vec{r}}_i \vec{q}_i ]
             \nonumber \\
& & - \int_0^{t_f} \mathrm{d}t\,  m_R \vec{q}(t)
        \left[ \ddot{\vec{r}}(t) +
       \omega_0^2 \int_0^{\infty} \mathrm{d}s \exp(-s) \vec{r}(t + \tau_0 s) \right]
         \nonumber \\
&& + \frac{\mathrm{i}}{4} \int_0^{t_f} \mathrm{d}t \int_0^{t_f} \mathrm{d}t' D_1(t-t')
              \dot{\vec{q}}(t)  \dot{\vec{q}}(t'),
\end{eqnarray}
or, equivalently,
\begin{eqnarray}
  \label{eq:105}
{\mathcal{A}}_{\mathrm{cl}}[\vec{r}, \vec{q}] & = &
         m_R [ \dot{\vec{r}}_f \vec{q}_f
               -\dot{\vec{r}}_i \vec{q}_i]
   + \frac{\mathrm{i}}{2} \int_0^{t_f} \mathrm{d}t \; \vec{q}(t)
           \frac{\mathrm{d}}{\mathrm{d}t}
       \int_0^{t_f} \mathrm{d}t' D_1(t-t') \dot{\vec{q}}(t')
         \nonumber \\
&& + \frac{\mathrm{i}}{4} \int_0^{t_f} \mathrm{d}t \int_0^{t_f} \mathrm{d}t' D_1(t-t')
              \dot{\vec{q}}(t)  \dot{\vec{q}}(t').
\end{eqnarray}
Eq.~(\ref{eq:100}) shows that the solution $\vec{r}(t)$ is, in
general, complex due to the coupling to $\vec{q}(t)$ via the noise
kernel $D_1(t-t')$. Consider the decomposition of $\vec{r}(t)$
into real and imaginary part,
\begin{equation}
 \label{eq:106}
 \vec{r}(t) = \vec{r}^{(1)}(t) + \mathrm{i}\vec{r}^{(2)}(t),
\end{equation}
where $\vec{r}^{(1)}$ is a solution of the real part of
Eq.~(\ref{eq:100}), while $\vec{r}^{(2)}$ solves its imaginary
part,
\begin{equation}
  \label{eq:107}
  m_R \left[
             \ddot{\vec{r}}^{(2)}(t) + \omega_0^2
            \int_0^{\infty} \mathrm{d}s \exp(-s)
            \vec{r}^{(2)}(t + \tau_0 s) \right] =
     - \frac{1}{2} \frac{\mathrm{d}}{\mathrm{d}t}
        \int_0^{t_f} \mathrm{d}t' D_1(t-t') \dot{\vec{q}}(t').
\end{equation}

We now demonstrate that, in order to determine the action along
the classical paths, it suffices to find the homogeneous solution
$\vec{r}^{(1)}$ and to insert it in the action functional
\cite{GRABERT}. In other words we have
\begin{equation}
  \label{eq:108}
  {\mathcal{A}}_{\mathrm{cl}}[\vec{r}^{(1)}, \vec{q}] =
 {\mathcal{A}}_{\mathrm{cl}}[\vec{r}, \vec{q}],
\end{equation}
where
\begin{eqnarray}
  \label{eq:109}
  {\mathcal{A}}_{\mathrm{cl}}[\vec{r}^{(1)}, \vec{q}] &=&
 m_R [\dot{\vec{r}}_f^{(1)} \vec{q}_f -\dot{\vec{r}}_i^{(1)} \vec{q}_i ]
        \nonumber \\
 && + \frac{\mathrm{i}}{4} \int_0^{t_f}\mathrm{d}t  \int_0^{t_f} \mathrm{d}t'
          D_1(t-t') \dot{\vec{q}}(t)  \dot{\vec{q}}(t').
\end{eqnarray}
To proof this statement we first deduce from Eq.~(\ref{eq:107})
that
\begin{eqnarray}
  \label{eq:110}
\lefteqn{  \frac{\mathrm{i}}{2} \int_0^{t_f} \mathrm{d}t \; \vec{q}(t) \frac{\mathrm{d}}{\mathrm{d}t}
           \int_0^{t_f} \mathrm{d}t'D_1(t-t') \dot{\vec{q}}(t')} \nonumber \\
& = & -\mathrm{i} m_R \int_0^{t_f} \mathrm{d}t \; \vec{q}(t)
       \left[ \ddot{\vec{r}}^{(2)}(t) + \omega_0^2
           \int_0^{\infty} \mathrm{d}s \exp(-s) \vec{r}^{(2)}(t +   \tau_0 s) \right]
            \nonumber \\
& = & -\mathrm{i} m_R
      [\dot{\vec{r}}_f^{(2)} \vec{q}_f -\dot{\vec{r}}_i^{(2)} \vec{q}_i ]
          \nonumber \\
&&
    - \mathrm{i} m_R \int_0^{t_f} \mathrm{d}t
          \left[ \vec{r}^{(2)}(t) \ddot{\vec{q}}(t)
             + \omega_0^2 \int_0^{\infty} \mathrm{d}s \exp(-s)
                  \vec{r}^{(2)}(t+\tau_0 s) \vec{q}(t) \right].
\end{eqnarray}
The term within the square brackets is seen to vanish if one
employs Eq.~(\ref{eq:101}) and the same arguments that were used
to derive the equation of motion from the variation (\ref{eq:103})
of the action functional. Furthermore, we made use of
$\vec{r}^{(2)}(0) =\vec{r}^{(2)}(t_f) = 0$ which means that the
real part $\vec{r}^{(1)}(t)$ of the solution satisfies the given
boundary conditions. Hence we find
\begin{equation}
  \label{eq:111}
   \frac{\mathrm{i}}{2} \int_0^{t_f} \mathrm{d}t \; \vec{q}(t) \frac{\mathrm{d}}{\mathrm{d}t}
           \int_0^{t_f} \mathrm{d}t'D_1(t-t') \dot{\vec{q}}(t') =
 -\mathrm{i} m_R
      [\dot{\vec{r}}_f^{(2)} \vec{q}_f -\dot{\vec{r}}_i^{(2)} \vec{q}_i ],
\end{equation}
from which we finally obtain with the help of (\ref{eq:105}),
\begin{eqnarray}
  \label{eq:112}
  {\mathcal{A}}_{\mathrm{cl}}[\vec{r}, \vec{q}] & = &
        m_R [ \dot{\vec{r}}_f \vec{q}_f -\dot{\vec{r}}_i \vec{q}_i ]
          - \mathrm{i} m_R [\dot{\vec{r}}_f^{(2)} \vec{q}_f -
                   \dot{\vec{r}}_i^{(2)} \vec{q}_i] \nonumber \\
 & & + \frac{\mathrm{i}}{4} \int_0^{t_f} \mathrm{d}t \int_0^{t_f} 
            \mathrm{d}t' D_1(t-t')
             \dot{\vec{q}}(t)   \dot{\vec{q}}(t') \nonumber \\
& = &  {\mathcal{A}}_{\mathrm{cl}}[\vec{r}^{(1)}, \vec{q}].
\end{eqnarray}
This completes the proof of the above statement.

Summarizing, the procedure to determine the propagator function
for the electron can now be given as follows. One first solves the
equations of motion
\begin{eqnarray}
  \label{eq:113}
\ddot{\vec{r}}(t) + \omega_0^2 \int_0^{\infty} \mathrm{d}s
        \exp(-s) \vec{r}(t+ \tau_0 s) & = & 0, \\
   \label{eq:114}
\ddot{\vec{q}}(t) + \omega_0^2 \int_0^{\infty} \mathrm{d}s
        \exp(-s) \vec{q}(t - \tau_0 s) & = & 0,
\end{eqnarray}
together with the boundary conditions (\ref{eq:115}). With the
help of these solutions one then evaluates the classical action,
\begin{equation}
  \label{eq:116}
 {\mathcal{A}}_{\mathrm{cl}}[\vec{r}, \vec{q}] =
 m_R [ \dot{\vec{r}}_f \vec{q}_f -\dot{\vec{r}}_i \vec{q}_i ]
 + \frac{\mathrm{i}}{4} \int_0^{t_f} \mathrm{d}t \int_0^{t_f} 
 \mathrm{d}t' D_1(t-t')
 \dot{\vec{q}}(t)   \dot{\vec{q}}(t'),
\end{equation}
which immediately yields the propagator function
\begin{eqnarray}
  \label{eq:117}
  J(\vec{r}_f, \vec{q}_f, t_f;\vec{r}_i, \vec{q}_i) & = &
 N \exp \left\{ \mathrm{i} {\mathcal{A}}_{\mathrm{cl}} [\vec{r}, \vec{q}] \right\}
    \nonumber \\
& = & N \exp \left\{ \mathrm{i} m_R
         (\dot{\vec{r}}_f \vec{q}_f -\dot{\vec{r}}_i \vec{q}_i)
      + \Gamma(\vec{q}_f, \vec{q}_i, t_f) \right\}.
\end{eqnarray}
Here, $N$ is a normalization factor which is determined from the
normalization condition
\begin{equation}
  \label{eq:118}
  \int \mathrm{d}^3r_f  J(\vec{r}_f, \vec{q}_f=0, t_f;\vec{r}_i, \vec{q}_i)
 = \delta(\vec{q}_i).
\end{equation}
The function $\Gamma(\vec{q}_f, \vec{q}_i, t_f)$ introduced in
Eq.~(\ref{eq:117}) will be referred to as the {\em{decoherence
function}}. It is given in terms of the noise kernel $D_1(t-t')$
as
\begin{equation}
  \label{eq:119}
 \Gamma(\vec{q}_f, \vec{q}_i, t_f) = - \frac{1}{4}
    \int_0^{t_f} \mathrm{d} t   \int_0^{t_f} \mathrm{d}t' D_1(t-t')
        \dot{\vec{q}}(t)  \dot{\vec{q}}(t').
\end{equation}
Explicitly we find with the help of Eq.~(\ref{eq:57}),
\begin{equation}
  \label{eq:120}
  \Gamma = - \frac{1}{4} \int_0^{t_f} \mathrm{d}t \int_0^{t_f} \mathrm{d}t'
              \int_0^{\infty} \mathrm{d} \omega J(\omega) \coth
                (\beta \omega/2) \cos \omega (t-t')
              \dot{\vec{q}}(t)  \dot{\vec{q}}(t').
\end{equation}
The double time-integral can be written as
\begin{equation}
 \label{eq:121}
 {\mathrm{Re}} \int_0^{t_f} \mathrm{d}t \int_0^{t_f} \mathrm{d}t'
 \exp [\mathrm{i} \omega(t-t')]  \dot{\vec{q}}(t)  \dot{\vec{q}}(t')
 = \left| \int_0^{t_f} \mathrm{d}t \exp(i \omega t)\dot{\vec{q}}(t)
 \right|^2.
\end{equation}
Hence, the decoherence function takes the form
\begin{equation}
  \label{eq:122}
 \Gamma(\vec{q}_f, \vec{q}_i, t_f)= - \frac{1}{4}
  \int_0^{\infty} \mathrm{d} \omega J(\omega) \coth(\beta \omega/2)
         \left| \vec{Q}(\omega) \right|^2,
\end{equation}
where we have introduced
\begin{equation} \label{QDEF}
 \vec{Q}(\omega) \equiv \int_0^{t_f} \mathrm{d}t \exp( \mathrm{i} \omega t) \dot{\vec{q}}(t).
\end{equation}

It can be seen from the above expressions that $\Gamma$ is a
non-positive function. The decoherence function will be
demonstrated below to describe the reduction of electron coherence
through the influence of the radiation field.

\section{Decoherence Through the Emission of Bremsstrahlung}
As an example we shall investigate in this section the most simple
case, namely that of a free electron coupled to the radiation
field. This case is of particular interest since it allows an
exact analytical determination of the decoherence function and
already yields a clear physical picture for the decoherence
mechanism. Having determined the decoherence function, we proceed
with an investigation of its influence on the propagation of
electronic wave packets.

\subsection{Determination of the Decoherence Function}
We set $\omega_0=0$ to describe the free electron. The equations
of motion (\ref{eq:113}) and (\ref{eq:114}) with the boundary
conditions (\ref{eq:115}) can easily be solved to yield
 \begin{equation}
   \label{eq:123}
   \vec{r}(t) = \vec{r}_i + \frac{\vec{r}_f - \vec{r}_i}{t_f}t,
\qquad \vec{q}(t) = \vec{q}_i + \frac{\vec{q}_f -
\vec{q}_i}{t_f}t.
 \end{equation}
Making use of Eq.~(\ref{eq:117}) and determining the normalization
factor from Eq.~(\ref{eq:118}) we thus get the propagator
function,
\begin{eqnarray}
  \label{eq:126}
\lefteqn{J(\vec{r}_f, \vec{q}_f, t_f; \vec{r}_i, \vec{q}_i) =} \nonumber \\
& &
\left( \frac{m_R}{2 \pi t_f} \right)^3
  \exp \left\{\mathrm{i} \frac{m_R}{t_f} (\vec{r}_f - \vec{r}_i)
           (\vec{q}_f - \vec{q}_i) + \Gamma(\vec{q}_f,\vec{q}_i,t_f)
       \right\}.
\end{eqnarray}
As must have been expected $J$ is invariant under space
translations since it depends only on the difference $\vec{r}_f -
\vec{r}_i$. Furthermore, one easily recognizes that the
contribution
\begin{equation}
  \label{eq:127}
  G(\vec{r}_f - \vec{r}_i, \vec{q}_f - \vec{q}_i, t_f)
  \equiv  \left( \frac{m_R}{2 \pi t_f} \right)^3
       \exp \left\{\mathrm{i} \frac{m_R}{t_f} (\vec{r}_f - \vec{r}_i)
           (\vec{q}_f - \vec{q}_i)
       \right\}
\end{equation}
is simply the propagator function for the density matrix of a free
electron with mass $m_R$ for a vanishing coupling to the radiation
field. We can thus write the electron density matrix as follows,
\begin{eqnarray}
  \label{eq:128}
  \rho_m(\vec{r}_f, \vec{q}_f, t_f) &=&
  \int \mathrm{d}^3r_i \int \mathrm{d}^3 q_i
  G(\vec{r}_f - \vec{r}_i, \vec{q}_f - \vec{q}_i, t_f) \nonumber \\
  &~& \times \exp\left\{ \Gamma(\vec{q}_f, \vec{q}_i,t_f) \right\}
 \rho_m(\vec{r}_i, \vec{q}_i,0),
\end{eqnarray}
which exhibits that the decoherence function $\Gamma$ describes
the influence of the radiation field on the electron motion.

We proceed with an explicit calculation of the decoherence
function. It follows from Eqs.~(\ref{QDEF}) and (\ref{eq:123})
that
\begin{equation}
  \label{eq:129}
\vec{Q}(\omega) = \int_0^{t_f} \mathrm{d}t \exp(\mathrm{i} \omega t)
           \frac{\vec{q}_f - \vec{q}_i}{t_f}
= \frac{\exp(\mathrm{i} \omega t_f) -1}{\mathrm{i} \omega} \vec{w},
\end{equation}
where
\begin{equation} \label{VDEF}
\vec{w} \equiv \frac{1}{t_f} ( \vec{q}_f - \vec{q}_i).
\end{equation}
Therefore, the decoherence function is found to be
\begin{equation}
  \label{eq:130}
  \Gamma = - \frac{e^2 \vec{w}^2}{6 \pi^2} \int_0^{\Omega} \mathrm{d} \omega
           \frac{1 - \cos \omega t_f}{\omega} \coth(\beta
           \omega/2),
\end{equation}
where we have used expression (\ref{eq:58}) for the spectral
density $J(\omega)$. The decoherence function may be decomposed
into a vacuum contribution $\Gamma_{\mathrm{vac}}$ and a thermal
contribution $\Gamma_{\mathrm{th}}$,
\begin{equation}
  \label{eq:131}
  \Gamma = \Gamma_{\mathrm{vac}} + \Gamma_{\mathrm{th}},
\end{equation}
where
\begin{equation}
  \label{eq:132}
  \Gamma_{\mathrm{vac}} = - \frac{e^2 \vec{w}^2}{6 \pi^2}
       \int_0^{\Omega} \mathrm{d} \omega \frac{1- \cos \omega t_f}{\omega}
\end{equation}
and
\begin{equation}
  \label{eq:133}
  \Gamma_{\mathrm{th}} = - \frac{e^2 \vec{w}^2}{6 \pi^2}
      \int_0^{\Omega} \mathrm{d} \omega \frac{1- \cos \omega t_f}{\omega}
     \left[ \coth(\beta \omega/2) -1  \right].
\end{equation}

The frequency integral appearing in the vacuum contribution can be
evaluated in the following way. Substituting $x=\omega t_f$ we get
\begin{equation}
  \label{eq:134}
   \int_0^{\Omega} \mathrm{d} \omega \frac{1- \cos \omega t_f}{\omega} =
       \int_0^{\Omega t_f} \mathrm{d}x \frac{1 - \cos x}{x}
 = \ln \Omega t_f + C + O\left( \frac{1}{\Omega t_f} \right),
\end{equation}
where $C \approx 0.577$ is Euler's constant \cite{GRADSHTEYN}. For
$\Omega t_f \gg 1$ we obtain asymptotically
\begin{equation}
  \label{eq:135}
  \Gamma_{\mathrm{vac}} \approx - \frac{e^2 \vec{w}^2}{6 \pi^2}
                \ln \Omega t_f =
         - \frac{e^2}{6 \pi^2}  \ln \Omega t_f
                \frac{(\vec{q}_f - \vec{q}_i)^2}{t_f^2}.
\end{equation}

To determine the thermal contribution $\Gamma_{\mathrm{th}}$ we
first write Eq.~(\ref{eq:133}) as follows,
\begin{equation}
  \label{eq:138}
  \Gamma_{\mathrm{th}} = - \frac{e^2 \vec{w}^2}{6 \pi^2}
      \int_0^{t_f} \mathrm{d}t \int_0^{\Omega} \mathrm{d} \omega
        \left[\coth (\beta \omega/2) -1 \right] \sin \omega t
       \equiv - \frac{e^2 \vec{w}^2}{6 \pi^2}  I.
\end{equation}
Introducing the integration variable $x=\beta \omega$ we can cast
the double integral $I$ into the form
\begin{displaymath}
  I = \frac{1}{\beta} \int_0^{t_f} \mathrm{d}t \int_0^{\beta \Omega} \mathrm{d}x
      \left[ \coth(x/2) -1 \right] \sin\left( tx/\beta \right).
\end{displaymath}
Here, we have $\beta \Omega = \hbar \Omega/k_B T$ and, using the
cutoff $\hbar \Omega = mc^2$, we get
\begin{displaymath}
  \beta \Omega = \frac{m c^2}{k_B T}.
\end{displaymath}
For temperatures $T$ obeying
\begin{equation}
  \label{eq:139}
  k_B T \ll mc^2
\end{equation}
the upper limit of the $x$-integral may be shifted from $\beta
\Omega$ to $\infty$. Condition (\ref{eq:139}) states that
\begin{displaymath}
  \frac{\hbar^2}{m k_B T} \gg \frac{\hbar^2}{m^2 c^2},
\end{displaymath}
which means that the thermal wavelength
$\bar{\lambda}_{\mathrm{th}}=\hbar/ \sqrt{2mk_B T}$ is much larger
than the Compton wavelength,
\begin{equation}
  \label{eq:140}
  \bar{\lambda}_{\mathrm{th}} \gg \bar{\lambda}_C.
\end{equation}
Thermal and Compton wavelength are of equal size at a temperature
of about $10^9$ Kelvin. Condition (\ref{eq:139}) therefore means
that $T \ll 10^9 \; \mathrm{K}$. Under this condition we now
obtain
\begin{eqnarray}
  \label{eq:141}
  I & \approx & \frac{1}{\beta} \int_0^{t_f} \mathrm{d}t
          \int_0^{\infty} \mathrm{d}x \left[ \coth(x/2) -1  \right]
           \sin\left( tx/\beta \right) \nonumber \\
   & =&  \frac{1}{\beta} \int_0^{t_f} \mathrm{d}t
            \left[ \pi \coth \left( \frac{\pi t}{\beta} \right)
               - \frac{\beta}{t} \right] \nonumber \nonumber \\
   & = & \ln \left( \frac{\sinh \left( \pi t_f / \beta \right)}{\pi t_f / \beta}
         \right),
\end{eqnarray}
where we have employed the formula
\begin{equation}
  \label{eq:136}
  \int_0^{\infty} \mathrm{d}x \left[ \coth(x/2) -1 \right] \sin \tau x
         = \pi \coth (\pi \tau) - \frac{1}{\tau}.
\end{equation}
The quantity
\begin{equation}
  \label{eq:144}
  \tau_B \equiv \frac{\beta}{\pi} = \frac{\hbar}{\pi k_B T} \approx 2.4 \cdot
  10^{-12} \; {\mathrm{s}}/{\mathrm{T[K]}}
\end{equation}
represents the correlation time of the thermal radiation field.
Putting these results together we get the following expression for
the thermal contribution to the decoherence function,
\begin{equation}
  \label{eq:155}
  \Gamma_{\mathrm{th}} \approx - \frac{e^2}{6 \pi^2}
           \ln \left( \frac{\sinh (t_f/\tau_B)}{t_f/\tau_B} \right)
   \frac{(\vec{q}_f - \vec{q}_i)^2}{t_f^2}.
\end{equation}
Adding this expression to the vacuum contribution (\ref{eq:135})
and introducing $\alpha = e^2/4 \pi \hbar c$ and further factors
of $c$, we can finally write the expression for the decoherence
function as
\begin{equation}
  \label{eq:157}
  \Gamma(\vec{q}_f, \vec{q}_i,t_f) \approx
    - \frac{2 \alpha}{3 \pi}
      \left[ \ln \Omega t_f +
              \ln \left(\frac{\sinh(t_f/\tau_B)}{t_f/\tau_B} \right)
                        \right]
          \frac{ (\vec{q}_f - \vec{q}_i)^2}{(c t_f)^2}.
\end{equation}
Alternatively, we may write
\begin{equation}
  \label{eq:163}
  \Gamma(\vec{q}_f, \vec{q}_i,t_f) = -
 \frac{(\vec{q}_f - \vec{q}_i)^2}{2 L(t_f)^2},
\end{equation}
where the quantity $L(t_f)$ defined by
\begin{equation}
  \label{eq:164}
  L(t_f)^2 \equiv \frac{3 \pi}{4 \alpha}
      \left[ \ln \Omega t_f
        + \ln \left( \frac{\sinh(t_f/\tau_B)}{t_f/\tau_B} \right)
        \right]^{-1} \cdot (c t_f)^2
\end{equation}
may be interpreted as a time-dependent {\em{coherence length}}.

The vacuum contribution $\Gamma_{\mathrm{vac}}$ to the decoherence
function (\ref{eq:157}) apparently diverges with the logarithm of
the cutoff $\Omega$. This is, however, an artificial
divergence which can be seen as follows. The decoherence function
is defined in terms of the Fourier transform $\vec{Q}(\omega)$ of
$\dot{\vec{q}}(t)$, see Eqs.~(\ref{eq:122}) and (\ref{QDEF}).
Evaluating $\vec{Q}(\omega)$ as in Eq.~(\ref{eq:129}) we assume
that the velocity is zero prior to the initial time $t=0$, that it
suddenly jumps to the value given by Eq.~(\ref{VDEF}), and that it
again jumps to zero at time $t_f$. This implies a force having the
shape of two $\delta$-function pulses around $t=0$ and $t=t_f$.
Such a force acts over two infinitely small time intervals and
leads to sharp edges in the classical path. More realistically one
has to consider a finite time scale $\tau_p$ for the action of the
force which must be still large compared to the radiation time
scale $\tau_0$. We may interpret the time scale $\tau_p$ as a
{\em{preparation time}} since it represents the time required to
prepare the initial state of a moving electron. A natural,
physical cutoff frequency of the order $\Omega \sim 1/\tau_p$ is
thus introduced by the preparation time scale $\tau_p$ and we may
set
\begin{equation}
  \label{eq:158}
  \Omega t_f = \frac{t_f}{\tau_p}
\end{equation}
in the following. It should be noted that the weak logarithmic
dependence on $\Omega$ shows that the precise value of the
preparation time scale $\tau_p$ is rather irrelevant. The
important point is that the preparation time introduces a new time
scale which removes the dependence on the cutoff. The vacuum
decoherence function can thus be written,
\begin{equation} \label{VACUUM-DECOHERENCE}
  \Gamma_{\mathrm{vac}} \approx
    - \frac{2 \alpha}{3 \pi} \ln \left( \frac{t_f}{\tau_p} \right)
    \frac{ (\vec{q}_f - \vec{q}_i)^2}{(c t_f)^2},
\end{equation}
showing that it vanishes for large times essentially as
$t_f^{-2}$.

The thermal contribution $\Gamma_{\mathrm{th}}$ is determined by
the thermal correlation time $\tau_B$. For $T \longrightarrow 0$
we have $\tau_B \longrightarrow \infty$, and this contribution
vanishes. For large times $t_f \gg \tau_B$ the thermal decoherence
function may be approximated by
\begin{equation}
  \label{eq:159}
 \Gamma_{\mathrm{th}} \approx - \frac{2 \alpha}{3 \pi}
 \frac{t_f}{\tau_B} \frac{(\vec{q}_f - \vec{q}_i)^2}{(c t_f)^2},
\end{equation}
which shows that $\Gamma_{\mathrm{th}}$ vanishes as $t_f^{-1}$.
Thus, for short times the vacuum contribution dominates, whereas
the thermal contribution is dominant for large times. Both
contributions $\Gamma_{\mathrm{vac}}$ and $\Gamma_{\mathrm{th}}$
are plotted separately in Fig.~1 which clearly shows the crossover
between the two regions of time.

Eq.~(\ref{eq:164}) implies that the vacuum coherence length is
roughly of the order
\begin{equation} \label{L-ESTIMATE}
L(t_f)_{\mathrm{vac}} \sim c \cdot t_f.
\end{equation}
To see this let us assume a typical preparation time scale of the
order $\tau_p \sim 10^{-21}$s. If we take $t_f$ to be of the order
of 1s we find that $\ln (t_f/\tau_p ) \sim 48$. In the rather
extreme case $t_f \sim 10^{17}$, which is of the order of the age
of the universe, we get $\ln (t_f/\tau_p ) \sim 87$. On using
$3\pi/4\alpha \approx 322$ and Eq.~(\ref{eq:164}) for $T=0$ one is
led to the estimate (\ref{L-ESTIMATE}).

\begin{figure}
\begin{center}
\includegraphics[width=0.8\textwidth]{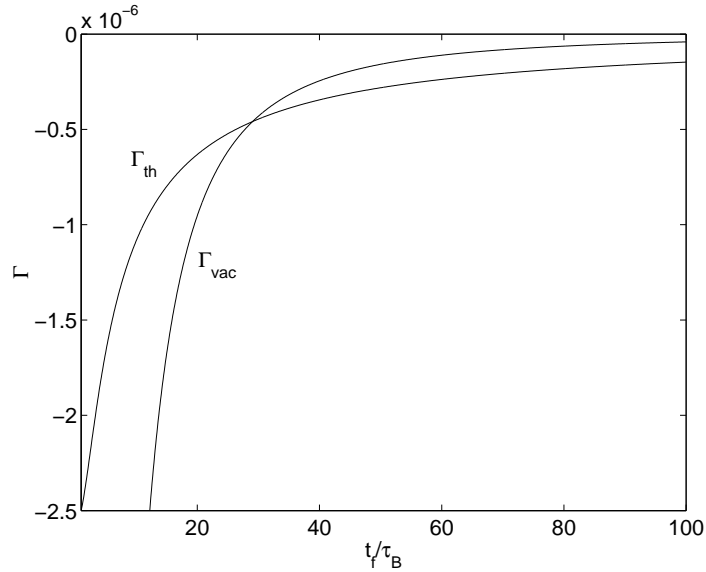}
\end{center}
\caption[]{The vacuum contribution $\Gamma_{\rm{vac}}$ and the
 thermal contribution $\Gamma_{\rm{th}}$ of the decoherence
 function $\Gamma$ (Eq.~(\ref{eq:157})). For a fixed value
 $|\vec{q}_f-\vec{q}_i| = 0.1 \cdot c\tau_B$,
 the two contributions are plotted against the time $t_f$
 which is measured in units of the thermal correlation time $\tau_B$.
 The temperature was chosen to be $T=1$K.
 One observes the decrease of both contributions for increasing time,
 demonstrating the vanishing of decoherence effects for long
 times. The thermal contribution $\Gamma_{\rm{th}}$ vanishes as
 $t_f^{-1}$, while the vacuum contribution
 $\Gamma_{\mathrm{vac}}$ decays essentially as $t_f^{-2}$,
 leading to a crossover between two regimes dominated by the
 vacuum and by the thermal contribution, respectively.}
\end{figure}

\subsection{Wave Packet Propagation}
Having obtained an expression for the decoherence function
$\Gamma$ we now proceed with a detailed discussion of its physical
significance. For this purpose it will be helpful to investigate
first how $\Gamma$ affects the time-evolution of an electronic
wave packet. We consider the initial wave function at time $t=0$,
\begin{equation}
  \label{eq:165}
  \psi_0(\vec{x}) = \left(\frac{1}{2 \pi \sigma_0^2} \right)^{3/4}
               \exp \left[- \frac{(\vec{x} - \vec{a})^2}{4 \sigma_0^2}
                    - \mathrm{i} \vec{k}_0 (\vec{x} - \vec{a}) \right],
\end{equation}
describing a Gaussian wave packet centered at $\vec{x} = \vec{a}$
with width $\sigma_0$. With the help of Eqs.~(\ref{eq:127}),
(\ref{eq:128}) and (\ref{eq:163}) we get the position space
probability density at the final time $t_f$,
\begin{eqnarray}
  \label{eq:166}
 \lefteqn{  \rho_m(\vec{r}_f, t_f)  \equiv  \rho_m(\vec{r}_f, \vec{q}_f =0, t_f) } \\
 & = & \int \mathrm{d}^3r_i \int \mathrm{d}^3 q_i
       \left( \frac{m_R}{2 \pi t_f} \right)^3
       \exp\left[ - \frac{\mathrm{i}m_R}{t_f} (\vec{r}_f - \vec{r}_i) \vec{q}_i
                  - \frac{\vec{q}_i^2}{2 L(t_f)^2} \right] \nonumber \\
 &~&   \qquad \qquad \qquad \times \psi_0(\vec{r}_i + \frac{1}{2} \vec{q}_i)
                     \psi_0^{\ast}(\vec{r}_i - \frac{1}{2} \vec{q}_i). \nonumber
\end{eqnarray}
The Gaussian integrals may easily be evaluated with the result,
\begin{equation}
  \label{eq:167}
  \rho_m(\vec{r}_f,t_f) = \left( \frac{1}{2 \pi \sigma(t_f)^2}\right)^{3/2}
              \exp\left[- \frac{(\vec{r}_f - \vec{b})^2}{2 \sigma(t_f)^2} \right],
\end{equation}
where
\begin{equation}
  \label{eq:168}
  \vec{b} \equiv \vec{a} - \frac{\vec{k}_0 t_f}{m_R}
\end{equation}
and
\begin{equation}
  \label{eq:169}
  \sigma(t_f)^2 \equiv \sigma_0^2 + \frac{t_f^2}{4 m_R^2 \sigma_0^2}
           + \frac{t_f^2}{m_R^2 L^2}.
\end{equation}
This shows that the wave packet propagates very much like that of
a free Schr\"odinger particle with physical mass $m_R$. The centre
$\vec{b}$ of the probability density moves with velocity
$-\vec{k}_0/m_R$, while its spreading, given by
Eq.~(\ref{eq:169}), is similar to the spreading
$\sigma(t_f)^2_{\mathrm{free}}$ which is obtained from the free
Schr\"odinger equation,
\begin{equation}
  \label{eq:170}
  \sigma(t_f)^2_{\mathrm{free}} = \sigma_0^2 +
         \frac{t_f^2}{4 m_R^2 \sigma_0^2}.
\end{equation}
If we write
\begin{equation}
  \label{eq:171}
  \sigma(t_f)^2 = \sigma_0^2 +  \frac{t_f^2}{4 m_R^2 \sigma_0^2}
             \left( 1 + \frac{4 \sigma_0^2}{L(t_f)^2} \right)
\end{equation}
we observe that the decoherence function affects the probability
density only though the width $\sigma(t_f)$ and leads to an
increase of the spreading. In view of the estimate
(\ref{L-ESTIMATE}) the correction term in Eq.~(\ref{eq:171}) is,
however, small for times satisfying
\begin{equation}
  \label{eq:172}
  L(t_f) \sim c \cdot t_f \gg \sigma_0.
\end{equation}
This means that the influence of the radiation field can safely be
neglected for times which are large compared to the time it takes
a light signal to travel the width of the wave packet.

\begin{figure}
\begin{center}
\includegraphics[width=0.8\textwidth]{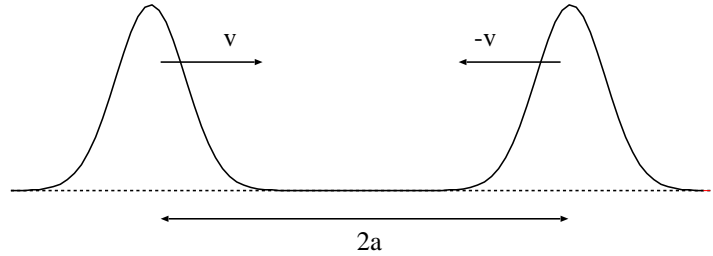}
\end{center}
\caption[]{Sketch of the interference experiment used to determine 
the decoherence factor. Two Gaussian wave packets with initial 
separation $2a$ approach each other with opposite 
velocities of equal magnitude $v=k_0/m_R$.}
\end{figure}

Let us now study the evolution of a superposition of two Gaussian
wave packets separated by a distance $2a$. This case has been
studied already by Barone and Caldeira \cite{CALDEIRA} who find,
however, a different result. We assume that the packets have equal
widths $\sigma_0$ and that they are centered initially at $\vec{x}
= \pm \vec{a} = \pm (a,0,0)$. The packets are supposed to approach
each other with the speed $v=k_0/m_R > 0$ (see Fig. 2). 
For simplicity the motion
is assumed to occur along the $x$-axis. Thus we have the initial
state
\begin{eqnarray}
  \label{eq:173}
  \psi_0(\vec{x}) & = & A_1 \left( \frac{1}{2 \pi \sigma_0^2} \right)^{3/4}
      \exp \left[ - \frac{(\vec{x} - \vec{a})^2}{4 \sigma_0^2}
          - \mathrm{i} \vec{k}_0 (\vec{x} - \vec{a}) \right]
                 \nonumber \\
 & + &  A_2 \left( \frac{1}{2 \pi \sigma_0^2} \right)^{3/4}
      \exp \left[ - \frac{(\vec{x} + \vec{a})^2}{4 \sigma_0^2}
          + \mathrm{i} \vec{k}_0 (\vec{x} + \vec{a}) \right],
\end{eqnarray}
where $\vec{k}_0 =(k_0,0,0)$ and $A_1$, $A_2$ are complex
amplitudes. Our aim is to determine the interference pattern that
arises in the moment of collision of the two packets at
$\vec{x}=0$. Using again Eqs.~(\ref{eq:127}), (\ref{eq:128}) and
(\ref{eq:163}) and doing the Gaussian integrals we find
\begin{eqnarray}
  \label{eq:174}
 \rho_m(\vec{r}_f, t_f) &=&
 \left( \frac{1}{2 \pi \sigma(t_f)^2} \right)^{3/2}
           \exp \left[ - \frac{\vec{r}_f^2}{2 \sigma(t_f)^2} \right]
 \nonumber \\
 &~& \times \left\{ |A_1|^2 + |A_2|^2 + 2 \mathrm{Re} A_1 A_2^{\ast}
           \exp[\varphi(\vec{r}_f)]\right\}.
\end{eqnarray}
We recognize a Gaussian envelope centered at $\vec{r}_f=0$ with
width $\sigma(t_f)$, an incoherent sum $|A_1|^2 + |A_2|^2$, and an
interference term proportional to $A_1 A_2^{\ast}$. The
interference term involves a complex phase given by
\begin{equation}
  \label{eq:175}
  \varphi(\vec{r}_f) = - 2 \mathrm{i} \vec{k}_0 \vec{r}_f (1- \varepsilon)
- \frac{2 a^2}{L(t_f)^2} (1- \varepsilon).
\end{equation}
The term $-2\mathrm{i}\vec{k}_0 \vec{r}_f$ describes the usual interference
pattern as it occurs for a free Schr\"odinger particle, while the
contribution $2\mathrm{i}\vec{k}_0 \vec{r}_f \varepsilon$ leads to a
modification of the period of the pattern. The final time $t_f = a
m_R/k_0$ is the collision time and
\begin{equation}
  \label{eq:177}
  v = \frac{a}{t_f} = \frac{k_0}{m_R}
\end{equation}
is the speed of the wave packets. The factor $\varepsilon$ is
given by
\begin{equation}
  \label{eq:178}
  \varepsilon \equiv \frac{t_f^2}{m_R^2 L(t_f)^2 \sigma(t_f)^2}
       = \left( 1 + \frac{L(t_f)^2}{4 \sigma_0^2} +
                      \frac{m_R^2 \sigma_0^2 L(t_f)^2}{t_f^2} \right)^{-1}.
\end{equation}
Obviously we always have $0 < \varepsilon <1$. Furthermore, for
the situation considered in Eq.~(\ref{eq:172}) we have $\epsilon
\ll 1$. Thus, we get
\begin{equation}
  \label{eq:179}
 \varphi(\vec{r}_f) = - 2 \mathrm{i} \vec{k}_0 \vec{r}_f - \frac{2 a^2}{L(t_f)^2}.
\end{equation}

The last expression clearly reveals that the real part of the
phase $\varphi(\vec{r}_f)$ describes decoherence, namely a
reduction of the interference contrast described by the factor
\begin{equation}
  \label{eq:180}
  D = \exp \left[- \frac{(2a)^2}{2L(t_f)^2} \right] =
  \exp \left[ - \frac{\mathrm{distance}^2}{2 ({\mbox{coherence length}})^2}
       \right],
\end{equation}
which multiplies the interference term. As was to be expected from
the general formula for $\Gamma$, the decoherence factor $D$ is
determined by the ratio of the distance of the two wave packets to
the coherence length.

Alternatively, we can write the decoherence factor in terms of the
velocity (\ref{eq:177}) of the wave packets. In the vacuum case we
then get
\begin{equation}
  \label{eq:181}
 D_{\mathrm{vac}} = \exp \left[- \frac{8 \alpha}{3 \pi} \ln
                    \left( \frac{t_f}{\tau_p} \right)
                    \left( \frac{v}{c} \right)^2 \right].
\end{equation}
This clearly demonstrates that it is the motion of the wave
packets which is responsible for the reduction of of the
interference contrast: If one sets into relative motion the two
components of the superposition in order to check locally their
capability to interfere, a decoherence effect is caused by the
creation of a radiation field. As can be seen from
Eq.~(\ref{eq:130}) the spectrum of the radiation field emitted
through the moving charge is proportional to $1/\omega$ which is a
typical signature for the emission of bremsstrahlung. Thus we
observe that the physical origin for the loss of coherence
described by the decoherence function is the creation of
bremsstrahlung.

It is important to recognize that the frequency integral of
Eq.~(\ref{eq:130}) converges for $\omega \rightarrow 0$, see
Eq.~(\ref{eq:134}). The decoherence function $\Gamma$ is thus
infrared convergent which is obviously due to the fact that we
consider here a process on a finite time scale $t_f$. This means
that we have a natural infrared cutoff of the order of
$\Omega_{\mathrm{min}} \sim 1/t_f$, in addition to the natural
ultraviolet cutoff $\Omega \sim 1/\tau_p$ introduced earlier. The
important conclusion is that the decoherence function is therefore
infrared as well as ultraviolet convergent.

It might be instructive, finally, to compare our results with the
corresponding expressions which are derived from the famous
Caldeira-Leggett master equation in the high-temperature limit
(see, e.g. \cite{GARDINER}). From the latter one finds the
following expression for the coherence length
\begin{equation}
  \label{eq:182}
 L(t_f)^2_{\mathrm{CL}} = \frac{\bar{\lambda}_{\mathrm{th}}^2}{2 \gamma t_f},
\end{equation}
where $\gamma$ is the relaxation rate. This is to be compared with
the expressions (\ref{eq:164}) for the coherence length. For large
temperatures we have the following dominant time and temperature
dependence,
\begin{equation}
  \label{eq:183}
 L(t_f)^2 \sim \frac{t_f}{T} \qquad \mathrm{and} \qquad
 L(t_f)^2_{\mathrm{CL}} \sim \frac{1}{T t_f}.
\end{equation}
Hence, while both expressions for the coherence length are
proportional to the inverse temperature, the time dependence is
completely different. Namely, for $t_f \longrightarrow \infty$ we
have
\begin{equation}
  \label{eq:184}
 L(t_f)^2 \longrightarrow \infty \qquad \mathrm{and} \qquad
 L(t_f)^2_{\mathrm{CL}} \longrightarrow 0,
\end{equation}
and, therefore, complete coherence in the case of bremsstrahlung
and total destruction of coherence in the Caldeira-Leggett case.

\section{The Harmonically Bound Electron in the Radiation Field}
As a further illustration let us investigate briefly the case of
an electron in the radiation field moving in a harmonic external
potential. Another approach to this problem may be found in
\cite{SPOHN}, where the authors arrive, however, at the conclusion
that there is no decoherence effect in the vacuum case.

We take $\omega_0> 0$ and solve the equation of motion
(\ref{eq:113}) with the help of the ansatz
\begin{equation}
  \label{eq:186}
  r(t) = r_0 \exp (zt),
\end{equation}
where, for simplicity, we consider the motion to be
one-dimensional. Substituting this ansatz into (\ref{eq:113}) one
is led to a cubic equation for $z$,
\begin{equation}
  \label{eq:190}
 z^2 - \tau_0 z^3 + \omega_0^2 =0.
\end{equation}
For vanishing coupling to the radiation field ($\tau_0 =0$) the
solutions are located at $z_{\pm} = \pm \mathrm{i} \omega_0$, describing
the free motion of a harmonic oscillator with frequency
$\omega_0$.

For $\tau_0 > 0$ the cubic equation has three roots, one is real
and the other two are complex conjugated to each other. The real
root corresponds to the runaway solution and must be discarded.
Let us assume that the period of the oscillator is large compared
to the radiation time,
\begin{equation}
  \label{eq:193}
  \tau_0 \ll \frac{1}{\omega_0}.
\end{equation}
Because of $\tau_0 \sim 10^{-24}$s this assumption is well
satisfied even in the regime of optical frequencies. We may thus
determine the complex roots to lowest order in $\omega_0 \tau_0$,
\begin{equation}
  \label{eq:194}
 z_{\pm} = \pm \mathrm{i} \omega_0 - \frac{1}{2} \tau_0 \omega_0^2.
\end{equation}
The purely imaginary roots $\pm \mathrm{i} \omega_0$ of the undisturbed
harmonic oscillator are thus shifted into the negative half plane
under the influence of the radiation field. The negative real part
describes the radiative damping. In fact, we see that $r(t)$
decays as $\exp(-\gamma t/2)$, where
\begin{equation}
  \label{eq:197}
  \gamma = \tau_0 \omega_0^2
         = \frac{2}{3} \alpha  \frac{\hbar \omega_0^2}{m_R c^2}
\end{equation}
is the damping constant for radiation damping \cite{JACKSON}. In
the following we consider times $t_f$ of the order of magnitude of
one period $\omega_0 t_f \sim 1$. Because of $\gamma t_f =
(\omega_0\tau_0) (\omega_0 t_f)$ we then have $\gamma t_f \sim
\tau_0 \omega_0 \ll 1$. In this case the damping can be neglected
and we may use the free solution in order to determine the
decoherence function.

Let us consider again the case of a superposition of two Gaussian
wave packets in the harmonic potential. The packets are initially
separated by a distance $2a$ and approach each other with opposite
velocities of equal magnitude such that they collide after a
quarter of a period, $t_f = \pi/2\omega_0$. The corresponding free
solution $q(t)$ is therefore given by
\begin{equation}
  \label{eq:200}
  q(t) = q_i \cos \omega_0 t +  q_f \sin \omega_0 t.
\end{equation}
To describe the situation we have in mind we take $q_i=2a$
(initial separation of the wave packets) and $q_f=0$ (to get the
probability density). Hence, we have
\begin{equation}
\dot{q}(t) = -2a\omega_0 \sin \omega_0 t,
\end{equation}
and we evaluate the Fourier transform,
\begin{eqnarray*}
  Q(\omega) & = & \int_0^{t_f} \mathrm{d}t \exp( \mathrm{i}\omega t) \dot{q}(t) \\
  & = & a\omega_0
  \left[ \frac{\exp(\mathrm{i} [\omega + \omega_0] t_f) -1}{\omega + \omega_0}
        -\frac{\exp(\mathrm{i} [\omega - \omega_0] t_f) -1}{\omega - \omega_0}
        \right].
\end{eqnarray*}
This yields the decoherence function
\begin{eqnarray}
  \label{eq:204}
\lefteqn{\Gamma \equiv \Gamma(q_f=0,q_i,t_f)} \\
& &= - \frac{e^2(a\omega_0)^2}{6 \pi^2}
 \int_0^{\Omega} \!\!\!\!\! \mathrm{d} \omega \omega \left[
 \frac{1 - \cos (\omega + \omega_0)t_f}{(\omega + \omega_0)^2}
+\frac{1 - \cos (\omega - \omega_0)t_f}{(\omega - \omega_0)^2}
 \right] \coth \left( \frac{\beta \omega}{2} \right). \nonumber
\end{eqnarray}

We discuss the case of zero temperature. The frequency integral in
Eq. (\ref{eq:204}) then approaches asymptotically the value $2\ln
\Omega t_f$ which leads to the following expression for the
decoherence factor,
\begin{equation}
  \label{eq:208}
  D_{\mathrm{vac}} = \exp \Gamma_{\mathrm{vac}} =
 \exp \left[- \frac{8 \alpha}{3 \pi} \ln \left( \frac{t_f}{\tau_p} \right)
      \left\langle \left( \frac{v}{c} \right)^2 \right\rangle \right].
\end{equation}
The interesting point to note here is that this equation is the
same as Eq.~(\ref{eq:181}) for the free electron, with the only
difference that the square $(v/c)^2$ of the velocity, which was
constant in the previous case, must now be replaced with its time
averaged value $\langle (v/c)^2 \rangle$.

\section{Destruction of Coherence of Many-Particle States}
For a single electron the vacuum decoherence factor (\ref{eq:181})
turns out to be very close to 1, as can be illustrated by means of
the following numerical example. We take $\tau_p$ to be of the
order of $10^{-21}$s and $t_f$ of the order of 1s. Using a
velocity $v$ which is already as large as $1/10$ of the speed of
light, one finds that $\Gamma_{\mathrm{vac}} \sim 10^{-2}$,
corresponding to a reduction of the interference contrast of about
$1\%$. This demonstrates that the electromagnetic field vacuum is
quite ineffective in destroying the coherence of single electrons.

For a superposition of many-particle states the above picture can
lead, however, to a dramatic increase of the decoherence effect.
Consider the superposition
\begin{equation} \label{SUPER-N}
|\psi\rangle = |\psi_1\rangle + |\psi_2\rangle
\end{equation}
of two well-localized, spatially separated $N$-particle states
$|\psi_1\rangle$ and $|\psi_2\rangle$. We have seen that
decoherence results from the imaginary part of the influence phase
functional $\Phi[J_c,J_a]$, that is from the last term on the
right-hand side of Eq.~(\ref{eq:39}) involving the anti-commutator
function $D_1(x-x')_{\mu\nu}$ of the electromagnetic field. Thus,
it is the functional
\begin{equation}
 \Gamma[J_c] = - \frac{1}{4} \int_{t_i}^{t_f} \mathrm{d}^4x \int_{t_i}^{t_f}
 \mathrm{d}^4x' D_1(x-x')_{\mu \nu} J_c^{\mu}(x) J_c^{\nu}(x'),
\end{equation}
which is responsible for decoherence. This shows that the
decoherence function for $N$-electron states scales with the
square $N^2$ of the particle number. Thus we conclude that for the
case of the superposition (\ref{SUPER-N}) the decoherence function
must be multiplied by a factor of $N^2$, that is the decoherence
factor for $N$-particle states takes the form,
\begin{equation}
  \label{DEC-N}
 D_{\mathrm{vac}}^N \sim \exp \left[- \frac{8 \alpha}{3 \pi} \ln
                    \left( \frac{t_f}{\tau_p} \right)
                    \left( \frac{v}{c} \right)^2 N^2 \right].
\end{equation}
This scaling with the particle number obviously leads to a
dramatic increase of decoherence for the superposition of
$N$-particle states. To give an example we take $N = 6 \cdot
10^{23}$, corresponding to 1 mol, and ask for the maximal velocity
$v$ leading to a $1\%$ suppression of interference. With the help
of (\ref{DEC-N}) we find that $v \sim 10^{-16}$m/s. This means
that, in order to perform an interference experiment with 1 mol
electrons with only $1\%$ decoherence, a velocity of at most
$10^{-16}$m/s may be used. For a distance of 1m this implies, for
example, that the experiment would take $3 \times 10^8$ years!

\section{Conclusions}
In this paper the equations governing a basic decoherence
mechanism occurring in QED have been developed, namely the
suppression of coherence through the emission of bremsstrahlung.
The latter is created whenever two spatially separated wave
packets of a coherent superposition are moved to one place, which
is indispensable if one intends to check locally their capability
to interfere. We have seen that the decoherence effect through the
electromagnetic radiation field is extremely small for single,
non-relativistic electrons. The decoherence mechanism is thus very
ineffective on the Compton length scale. An important conclusion
is that decoherence does {\em{not}} lead to a localization of the
particle on arbitrarily small length scales and that no problems
with associated UV-divergences arise here.

The decoherence mechanism through bremsstrahlung exhibits a highly
non-Markovian character. As a result the usual picture of
decoherence as a decay of the off-diagonals in the reduced density
matrix does not apply. In fact, consider a superposition of two
wave packets with zero velocity. The expression (\ref{eq:163}) for
the decoherence function together with the estimate
$L(t_f)_{\mathrm{vac}} \sim c \cdot t_f$ for the vacuum coherence
length $L(t_f)_{\mathrm{vac}}$ show that decoherence effects are
negligible for times $t_f$ which are large in comparison to the
time it takes light to travel the distance between the wave
packets. The off-diagonal terms of the reduced density matrix for
the electron do therefore not decay at all, which shows the
profound difference between the decoherence mechanism through
bremsstrahlung and other decoherence mechanisms (see, e.~g.
\cite{JOOS}).

A result of particular interest from a fundamental point of view
is that coherence can already be destroyed by the presence of the
electromagnetic field vacuum if superpositions of many-particle
states are considered. An important conclusion which can be drawn
from this picture of decoherence in QED refers to various
alternative approaches to decoherence and the closely related
measurement problem of quantum mechanics: In recent years several
attempts have been made to modify the Schr\"odinger equation by
the addition of stochastic terms with the aim to explain the
non-existence of macroscopic superpositions through some kind of
macrorealism. Namely, the random terms in the Schr\"odinger
equation lead to a spontaneous destruction of superpositions in
such a way that macroscopic objects are practically always in
definite localized states. Such approaches obviously require the
introduction of previously unknown physical constants. In the
stochastic theory of Ghirardi, Pearle and Rimini
\cite{GHIRARDI90a}, for example, a single particle microscopic
jump rate of about $10^{-16}\rm{s}^{-1}$ has to be introduced such
that decoherence is extremely weak for single particles but acts
sufficiently strong for many particle assemblies. It is
interesting to observe that the decoherence effect caused by the
presence of the quantum field vacuum yields a similar time scale
in a completely natural way without the introduction of new
physical parameters. Thus, QED indeed provides a consistent
picture of decoherence and it seems unnecessary to propose new ad
hoc theories for this purpose.

It must be emphasized that the above picture of decoherence in QED
has been derived from the well-established basic postulates of
quantum mechanics and quantum field theory. It therefore does not,
of course, constitute a logical disprove of alternative
approaches. However, it does represent an example for a basic
decoherence mechanism in a microscopic quantum field theory. In
particular, it provides a unified explanation of decoherence which
does not suffer from problems with renormalization (as they occur,
e.g. in alternative theories \cite{GHIRARDI90b}) and which does
not exclude a priori the existence of macroscopic quantum
coherence. Only under certain well-defined conditions regarding
time scales, relative velocities and the structure of the state
vector, it is true that decoherence becomes important. Thus,
decoherence is traced back to a dynamical effect and not to a
modification of the basic principles of quantum mechanics.

In this paper we have discussed in detail only the
non-relativistic approximation of the reduced electron dynamics.
For a treatment of the full relativistic theory, including a
Lorentz invariant characterization of the decoherence induced by
the vacuum field, one can start from the formal development given
in section 2. An investigation along these lines could also be of
great interest for the study of measurement processes in the
relativistic domain \cite{BREUER99}.

\end{document}